\documentclass[prd,preprint,superscriptaddress,amsmath,amssymb,nofootinbib]{revtex4}
\usepackage{graphicx}% Include figure files
\usepackage{dcolumn}% Align table columns on decimal point
\usepackage{bm}% bold math
\usepackage{amssymb}
\usepackage{amsmath}
\usepackage{epsfig}    
\usepackage{color}
\usepackage{slashed}
\usepackage{hhline}
\usepackage{bbm}
\usepackage{tikz-feynman}
\tikzfeynmanset{compat=1.1.0}

\def\be{\begin{equation}}
\def\ee{\end{equation}}
\newcommand{\bea}{\begin{eqnarray}}
\newcommand{\eea}{\end{eqnarray}}
\newcommand{\nn}{\nonumber}

\begin{document}

%%%%%%%%%
%\title{Self-Consistent Framework for Neutrino Masses: Cut-off Scale from RGE Flow and Loop-Induced VEVs under Non-Invertible Symmetry  }
\title{Dirac one-loop seesaw in a non-invertible fusion rule}

\author{Hiroshi Okada}
\email{hiroshi3okada@htu.edu.cn}
\affiliation{Department of Physics, Henan Normal University, Xinxiang 453007, China}

\author{Labh Singh}
\email{sainilabh5@gmail.com}
\affiliation{Department of Physics and Astronomical Science, Central University of Himachal Pradesh, Dharamshala,
Himachal Pradesh 176215, India}

\date{\today}

\begin{abstract}
\noindent We propose a radiative Dirac neutrino mass model stabilized by a non-invertible fusion rule originating from a $Z_3 \times Z_3'$ gauging. The imposed symmetry forbids tree‑level Yukawa couplings and ensures that neutrino masses are generated only at the one‑loop level through the exchange of exotic fermions and inert scalars. This minimal framework simultaneously accommodates neutrino masses and mixings consistent with current oscillation data, while providing a viable dark matter candidate. We analyze lepton flavor violating processes and lepton anomalous magnetic moments, finding that all contributions remain well below present experimental bounds. In the dark matter sector, the bosonic singlet emerges as a promising candidate with relic density compatible with cosmological observations, whereas the fermionic option is strongly disfavored due to suppressed annihilation cross sections. Our study demonstrates that non‑invertible fusion rules can serve as a powerful organizing principle for constructing minimal and phenomenologically consistent extensions of the Standard Model, linking neutrino physics and dark matter within a unified radiative framework.
\end{abstract}

\maketitle

\section{Introduction}
\noindent The origin of neutrino masses and the nature of dark matter (DM) are two of the most significant pieces of evidence for physics beyond the Standard Model (SM). While Majorana neutrinos are often discussed in the context of the seesaw mechanism, Dirac neutrinos offer an equally compelling framework. 
In fact, the Dirac versus Majorana nature of neutrinos remains experimentally unresolved, 
particularly, the non-observation of neutrinoless double beta decay ($0\nu\beta\beta$) plays a crucial role, as dictated by the Schechter--Valle black box theorem \cite{Schechter:1981bd}. The theorem states that the observation of $0\nu\beta\beta$ would unambiguously imply the existence of a Majorana mass term for neutrinos, independent of the underlying mechanism. 
Consequently, the continued null results in $0\nu\beta\beta$ searches provide a strong motivation to consider Dirac type-neutrino scenario.
\\
%%%
In conventional Dirac-type scenarios~\cite{CentellesChulia:2016fxr, CentellesChulia:2017koy, Bonilla:2018ynb, CentellesChulia:2018bkz, CentellesChulia:2019xky, Peinado:2019mrn, Han:2018zcn, Wang:2017mcy, Borah:2017leo, Jana:2019mgj, Jana:2019mez, Calle:2019mxn, Nanda:2019nqy, Ma:2019byo, Ma:2019iwj, Correia:2019vbn, Saad:2019bqf, Ma:2019yfo, Guo:2020qin, delaVega:2020jcp, Borgohain:2020csn, Leite:2020wjl, Chulia:2021jgv, Bernal:2021ppq, Mishra:2021ilq, Biswas:2021kio, Mahanta:2021plx, Hazarika:2022tlc, CentellesChulia:2022vpz, Berbig:2022nre, Maharathy:2022gki, Chowdhury:2022jde, Biswas:2022vkq, Berbig:2022hsm, Mahapatra:2023oyh, Borah:2024gql, CentellesChulia:2024iom, Singh:2024imk, Kumar:2025cte, Batra:2025gzy, Kang:2026osw, Dasgupta:2021ggp, Okada:2014vla}, first discussed in Ref. \cite{Gu:2006dc}, the imposition of lepton number conservation or an equivalent symmetry is typically required.
This necessity arises from forbidding the Majorana mass term,
and symmetries such as continuous $U(1)$ or discrete groups have often been employed for this purpose.
Recently, however, a novel class of symmetries based on non-invertible fusion rules has been developed  and begun to find applications in phenomenology~\cite{Xu:2026nwh, Okada:2026bpp, Nomura:2026hli, Qu:2026omn, Kobayashi:2025ldi, Kobayashi:2024cvp, Kobayashi:2025znw,
Nomura:2025yoa, Suzuki:2025oov, Kobayashi:2025cwx, Nomura:2025sod,  Okada:2025kfm, Jangid:2025krp,
Jangid:2025thp, Nomura:2025tvz, Okada:2025adm, Okada:2026gxl, Chen:2025awz, Okada:2026iob, Liang:2025dkm,Kobayashi:2025thd, Kobayashi:2025rpx, Nakai:2025thw, Kobayashi:2025lar, Kobayashi:2025wty, Heckman:2024obe,Kaidi:2024wio,Funakoshi:2024uvy}.
 Unlike conventional group-based symmetries, these non-invertible structures possess distinctive features that go beyond group-theoretic framework.
In particular, they can universally preserve invariance only at the tree-level, with no guarantee of  stability once loop corrections are taken into account.
Exploiting this feature reduces the number of additional particles required and enables the construction of  more minimal loop-level models.\\
%%%
On the other hand,  sub-eV scale of neutrino mass suggests that it is unlikely to the other three fermion sectors in the SM, and we expect some different mechanisms to suppress the neutrino masses.
One of the attractive scenarios is that  the neutrino masses are generated at loop-level.
This can be understood that the neutrino does not directly interact with the SM Higgs but indirectly couples to the SM Higgs $via$ new particles.
By identifying a new particle as the DM candidate, we can construct a novel model that explains the tiny neutrino mass and DM simultaneously.
Since the DM is also neutral under the electric charge and is not directly detected yet, it would be natural to communicate the DM with the active neutrinos.

\noindent In this paper, we propose a Dirac loop-seesaw mechanism by introducing vector-like neutral fermions $\psi_{L,R}$ together with an inert boson sector consisting of an isospin singlet $S$ and doublet $\eta\equiv [\eta^+,\eta_0]^T$. These new particles generate one-loop Dirac neutrino masses and simultaneously provide a viable DM candidate.
To forbid direct tree-level interactions among the active neutrinos $L_L\equiv [\nu_L,\ell_L]^T$, the right-handed neutrinos $N_R$, and the SM Higgs  $H\equiv [h^+,(v+h+i z)/\sqrt2]^T$,
we impose a non-invertible fusion rule, specifically $Z_3$ gauging of $Z_3\times Z'_3$, which is one of the minimal realizations to our model.
{Following the phenomenological framework adopted in recent studies, we treat the fusion rule symmetry as an effective tree-level organizing principle whose associated operators can be generated radiatively at the loop level,} thereby inducing the Dirac neutrino term $\overline{L_L} H N_R$ as desired.
This framework provides a minimal setup for one-loop Dirac neutrino masses, as can be seen in our main text.
We then investigate the associated phenomenology, including lepton flavor violations (LFVs), the lepton anomalous magnetic dipole moment (lepton $g-2$), and DM feature, in addition to the neutrino masses and mixing patterns, through a detailed numerical analysis.
{We emphasize that the renormalization properties of four-dimensional quantum field theories with non-invertible symmetries are not yet completely understood and remain under active investigation. In particular, the treatment of loop-induced divergences and the associated counterterms for operators forbidden at tree level has been investigated in recent recent studies, including spurion-based approaches~\cite{Suzuki:2025bxg, Suzuki:2025kxz}, although a complete understanding has not yet been achieved. In this work, we adopt a phenomenological viewpoint and regard the non-invertible fusion rule as an effective organizing principle for constructing radiative Dirac neutrino mass models. Accordingly, our analysis should be interpreted as an effective field theory framework for exploring the phenomenological implications of non-invertible symmetries, rather than as a claim of a fully established ultraviolet-complete framework.}
In particular, we assume that higher-loop corrections to the neutrino mass are suppressed by additional loop factors and small couplings, as in standard perturbative estimates. However, we note that this assumption has not been rigorously justified in the presence of non-invertible symmetries, and a systematic treatment of higher-loop effects remains an important open problem for future work.

\noindent The remainder of this paper is organized as follows. 
In Sec. II we present the Dirac neutrino mass model setup based on the non‑invertible fusion rule and describe the particle content and interactions. 
Sec. III discusses the phenomenological implications, including lepton flavor violation, lepton $g-2$, and DM scenarios. 
Sec. IV provides the numerical analysis consistent with neutrino oscillation and cosmological data. 
Finally, Sec. V offers concluding remarks and discussion.

\section{The Model setup}
\label{sec:set}

\noindent In this section, we review how to construct the one-loop Dirac seesaw model with new particles imposing the fusion rule of $Z_3$ gauging of $Z_3 \times Z_3'$~\cite{Dong:2025jra}.

\subsection{$Z_3$ gauging of $Z_3 \times Z_3'$ fusion rule}
\label{subsec:fr}
\noindent Before discussing our model, we explain why  the $Z_3$ gauging of $Z_3 \times Z_3'$ fusion rule provides a minimal realization and  review its mathematical properties.
For convenience,  we hereafter adopt the short-hand notation $FR$ to denote the $Z_3$ gauging of $Z_3 \times Z_3'$ fusion rule , unless its detailed features are explicitly discussed.
In our framework, we introduce two types of $U(1)_Y$ singlet neutral fermions $\psi_R$ and $N_R$, which share the same chirality.
Since both these particles carry identical quantum numbers under the SM gauge symmetries, it is necessary to  distinguish them by
assigning different generators under the $FR$.
Furthermore, to preserve the Dirac nature of neutrinos, Majorana mass terms such as $\overline{\psi_R^C} \psi_R$ and  $\overline{N_R^C} N_R$  has to be forbidden.
This requirement implies that any $Z_2$ gauging fusion rules are disfavored, as they possess the feature $ {\mathbbm I}\in g\otimes g$, which inevitably allows Majorana mass terms. Here, $g$ denotes a generator belonging to a  $Z_2$ gauging fusion rule.
Consequently, we need to employ fusion rules that satisfy 
\[
{\mathbbm I}\notin g\otimes g,\quad {\mathbbm I}\notin g'\otimes g', \quad {\mathbbm I}\notin g\otimes g' , 
\] where $g(g')$ are assigned to be $\psi_R$ or $N_R$.
Moreover, $g(g')$ has to satisfy the condition ${\mathbbm I}\in g\otimes g^*\ (g'\otimes g'^*)$ in order to allow the construction of the  Dirac mass term $\overline{\psi_L} \psi_R$.
Taking all these requirements into account, one finds that the $Z_3$ gauging of $Z_3 \times Z_3'$ fusion rule provides one of the most minimal symmetry realizations.
\footnote{Equivalently, there is another possibility to adopt $Z_3$ gauging of $Z_9$ fusion rule~\cite{Dong:2025jra}. }

$Z_3$ gauging of $Z_3 \times Z_3'$ fusion rule consists of the following generators  $\{1, a, a^*, b, b^*\}$,
their multiplication rules are as follows:
\begin{align}
&    a \otimes a = a^*, \quad a^* \otimes a^* = a, \quad a \otimes a^* = {\mathbbm I}, \\
&    b \otimes b = b^*, \quad b^* \otimes b^* = b, \quad b \otimes b^* = {\mathbbm I} \oplus a \oplus a^*, \\
&   a \otimes b = b, \quad a^* \otimes b = b,\quad   a \otimes b^* = b^*, 
\quad a^* \otimes b^* = b^*.
\end{align}

\subsection{Model setup}
\label{subsec:model}

\begin{table}[t!]
\begin{tabular}{|c||c|c|c|c|c||c|c|c|}\hline\hline  
& ~$L_L$~& ~$\ell_R$~ & ~$\psi_{L}$~ & ~${\psi_R}$~ & ~${N_R}$~ & ~$H$~  & ~{$\eta$}~ & ~{$S$}~    \\\hline\hline 
%%%
$SU(2)_L$   & $\bm{2}$  & $\bm{1}$  & $\bm{1}$ & $\bm{1}$ & $\bm{1}$   & $\bm{2}$      & $\bm{2}$ & $\bm{1}$  \\\hline 
$U(1)_Y$    & $-\frac12$  & $-1$  & $0$  & $0$    & $0$ & $\frac12$    & $\frac12$   & $0$    \\\hline
FR   & $a^*$  & $a^*$ & $ b$ & $b$ & $a $  & $\mathbbm{I}$ & $b$& $b$ \\\hline 
 \end{tabular}
\caption{Charge assignments of the fields under the $SU(2)_L \times U(1)_Y$ gauge symmetry and our FR; $Z_3$ gauging of $Z_3\times Z_3'$. 
}\label{tab:1}
\end{table}

\noindent Our model set up is as follows.
{We introduce three generations of vector-like neutral fermions $\psi_{L,R}$ and right-handed neutral fermions $N_R$, together with an inert scalar sector consisting of an electroweak singlet scalar $S$ and an inert scalar doublet $\eta \equiv (\eta^+,\eta^0)^T$. Neither $S$ nor $\eta$ acquires a vacuum expectation value. These inert scalars participate in the one-loop generation of Dirac neutrino masses and can also furnish a dark matter candidate.}
New particles of ($\psi_{L,R}$, $N_R$, $S$, $\eta$) are respectively assigned by ($b,\ a,\ b,\ b$) under the FR. 
In addition, the SM lepton doublets $L_L$ and the charged-lepton singlets $\ell_R$ are assigned by $a^*$.
We summarize our particle contents and their assignments in Tab.~\ref{tab:1}.
Then,  the relevant Yukawa Lagrangian for the lepton sector is given by
\begin{align}
 -{\cal L} = y^\ell_i \bar{L}_{L_i} H \ell_{R_i} + f_{i\alpha} \bar{L}_{L_i} \tilde{\eta} \psi_{R_\alpha} + g_{\alpha a} \overline{\psi_{L_\alpha}} N_{R_a} S 
   +  M_{\psi_{\alpha}}  \overline{\psi_{L_\alpha}}  \psi_{L_\alpha} + \text{h.c.}, \label{eq:lagrangian}
\end{align}
where $\tilde \eta\equiv i\sigma_2 \eta^*$, $\sigma_2$ being the second Pauli matrix.
Without loss of generality, here, $y^\ell,\ M_\psi$ are diagonal basis.
We note that $\overline{\psi_{L,R}^C}\psi_{L,R} S$ is also allowed under our assignments, but this term does not affect our model. 
We list forbidden terms from the FR below:
\begin{align}
& \overline{L_L} \tilde{H} N_R,\quad  \overline{L_L} \tilde{H} \psi_R,\quad  \overline{L_L} \tilde{H} \psi_L^C,\quad
\overline{N_R^C} \psi_R,\quad \overline{\psi_R^C} \psi_R,\quad \overline{\psi_L^C} \psi_L,\quad \overline{N_R^C} N_R,\quad
   \overline{N_R^C} N_R S^{(*)}, \\
& H^\dag \eta,\quad  (H^\dag \eta)^2,\quad (H^\dag \eta) S,\quad  S^2,\quad  (S^*)^2.
   \label{eq:forbid}
\end{align}

% The tree-level Dirac mass term $\bar{L}_L \tilde{\Phi} N_R$ is forbidden by the FR[cite: 7].

\subsection{Higgs potential}
\label{sub:hp}
\noindent Our relevant Higgs potential is given by
\begin{align}
V &= -\mu_H^2 |H|^2 + \mu_S^2 |S|^2+ \mu_\eta^2 |\eta|^2 + \kappa (S^3+ \text{c.c.}) + \mu (H^\dag \eta S^*+ \text{c.c.})+\lambda_0 (H^\dag \eta S^2+\text{c.c.})\nn\\
&
+\lambda_H |H|^4
+\lambda_S |S|^4
+\lambda_\eta |\eta|^4
+\lambda_{HS} |H|^2|S|^2
+\lambda_{H\eta} |H|^2|\eta|^2
+\lambda'_{H\eta} |H^\dag \eta|^2
+\lambda_{S\eta} |S|^2|\eta|^2.
\label{eq:pot}
\end{align}
Instead of full analysis of the Higgs potential, here, we simply extract our relevant term $\mu$ that associates with the Dirac neutrino mass terms.
{Our model requires mixing between the singlet scalar $S$ and the neutral component of the inert doublet, $\eta^0$. This mixing is induced by the trilinear interaction
$\mu(H^\dagger \eta S^\ast + {\rm h.c.})$ after electroweak symmetry breaking and plays a crucial role in generating the one-loop Dirac neutrino masses.} Then, we define the mass eigenvectors $H_1$ and $H_2$ in terms of flavor eigenvectors $S$ and $\eta_0$ and its mixing $\theta$;
\begin{align}
S = c_\theta H_1 + s_\theta H_2,\quad 
\eta_0 = -s_\theta H_1 + c_\theta H_2,
\end{align}
where $s_\theta\sim \mu v/(m_1^2-m_2^2)$, $m_{1,2}$ are mass eigenvalues of $H_{1,2}$.

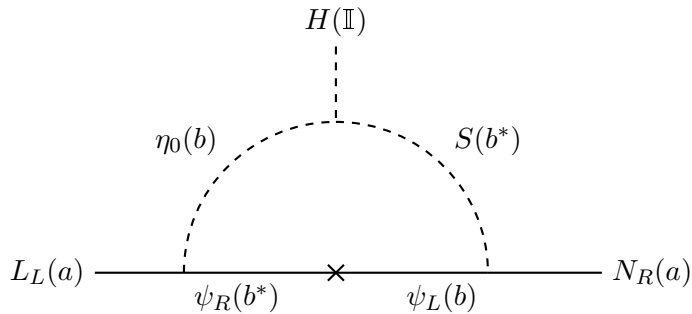
\begin{figure}[t]
\begin{center}
\begin{tikzpicture}
\begin{feynman}[large]
%%% definition of each vertex (and auxiliary points) %%%
\vertex (a1) {\(L_L({a})\)};
\vertex [right=1.8cm of a1] (b1);
\vertex [right=4.0cm of b1] (c1);
\vertex [right=1.5cm of c1] (d1) {\(N_R(a)\)};
\vertex [right=2.0cm of b1] (m1);
\vertex [above=2.0cm of m1] (e1);
\vertex [above=1.0cm of e1] (v1) {\(H(\mathbbm{I})\)};
\diagram [medium] {
%%% define all lines and labels on propagators %%%
(a1) -- (b1) -- [insertion={[size=3pt]0.5}, edge label'=\(\psi_R(b^*)~~~~~~~~~~~~~\psi_L(b)\)] (c1) -- (d1),
(b1) -- [scalar, quarter left, edge label=\(\eta_0(b)\)] (e1) -- [scalar, quarter left, edge label=\(S(b^*)\)] (c1),
(v1) -- [scalar] (e1),
};
\end{feynman}
\end{tikzpicture}
\end{center}
\vspace{-0.7cm}
\caption{One loop-induced Dirac neutrino masses. All vertices are invariant under FR. But one clearly finds that FR violates at one-loop, since $\mathbbm{1}\otimes a\otimes a = a^*\neq \mathbbm{1}$. 
% Note that all vertices and its one-loop are invariant under TY.
}
\label{fig:d1lp}
\end{figure}

\subsection{One-loop Dirac Neutrino Mass Generation}
\label{sub:neut}
\noindent As shown in the one-loop diagram in Fig.~\ref{fig:d1lp}, the Dirac mass is induced $via$ the mixing of $S$ and $H$.
The resultant Dirac neutrino mass matrix %$(m_\nu)_{ia}$ 
is given by
\begin{align}
(m_\nu)_{ia} \simeq \frac{s_\theta c_\theta}{(4\pi)^2} \sum_{\alpha} f_{i\alpha} F(m_1, m_2, M_{\psi_\alpha})  g_{a\alpha} ,
\end{align}
where the loop function $F(m_1, m_2, M_{\psi_\alpha})$ is defined as:
\begin{equation}
    F(m_1, m_2, M_{\psi_\alpha})=  M_{\psi_\alpha} \left[ \frac{m_1^2 \ln(m_1^2/M_{\psi_\alpha}^2)}{m_1^2 - M_{\psi_\alpha}^2} - \frac{m_2^2 \ln(m_2^2/M_{\psi_\alpha}^2)}{m_2^2 - M_{\psi_\alpha}^2} \right].
\end{equation}
For our convenience to a numerical analysis, we redefine the dimensionless neutrino mass matrix $\tilde m_\nu$ by extracting overall parameters as follows:
\begin{align}
&(\tilde m_\nu)_{ia} \equiv  \kappa_\nu   \sum_{\alpha} \tilde f_{i\alpha} \tilde  F(\tilde m_1, \tilde  m_2, \tilde M_{\psi_\alpha}) \tilde g_{a\alpha},\\
& \tilde  F(\tilde m_1, \tilde  m_2, \tilde M_{\psi_\alpha})
\equiv 
\tilde M_\alpha \left[\frac{\tilde m_1^2 \ln(\tilde m_1^2/\tilde M_{\psi_\alpha}^2)}{\tilde m_1^2 - \tilde M_{\psi_\alpha}^2} - \frac{\tilde m_2^2 \ln(\tilde m_2^2/\tilde M_{\psi_\alpha}^2)}{\tilde m_2^2 - \tilde M_{\psi_\alpha}^2} \right],\\
&\kappa_\nu\equiv \frac{s_\theta c_\theta M_{\psi_3} f_{33} g_{33}}{(4\pi)^2}, \label{eq:knu}
\end{align}
where $f\equiv f_{33} \tilde f$, $g\equiv g_{33} \tilde g$, $M_{\psi}=M_{\psi_3} \tilde M_\psi$, $\tilde m_{1,2}\equiv m_{1,2}/M_{\psi_3}$,
$f_{33},\ g_{33},\ M_{\psi_3}$ are respectively 3-3 components of $f,\ g,\ M_{\psi}$.
 From Eq.~(\ref{eq:knu}), $f_{33}$ is not a independent parameter whose  form is given by
 \begin{align}
 f_{33} = \frac{(4\pi)^2}{s_\theta c_\theta M_{\psi_3} g_{33}} \kappa_\nu. 
 \end{align}
 In our numerical analysis, we adopt the above form instead of $f_{33}$.

\noindent The neutrino mass eigenvalues $D_\nu\equiv \kappa_\nu \tilde D_\nu$ is diagonalized by $\tilde D_\nu = V^\dag_L \tilde m_\nu V_R$.
Since the charged-lepton mass matrix is diagonal basis, the observed mixing matrix is identical to $V_L$, which is denoted by $U_{\rm PMNS}$.
Then, $\kappa_\nu$ can be rewritten in terms of  the solar mass squared difference $\Delta m^2_{\rm sol}$ and dimensionless neutrino mass eigenvalues as follows:
\begin{align}
\kappa_\nu^2 \equiv \frac{\Delta m^2_{\rm sol}}{\tilde D_{\nu_2}^2-\tilde D_{\nu_1}^2}.
\end{align}
In this case, the atmospheric  mass squared difference $\Delta m^2_{\rm atm}$ is given by
\begin{align}
&({\rm NH}): \ \Delta m^2_{\rm atm}\equiv  \Delta m^2_{\rm sol}   \frac{\tilde D_{\nu_3}^2-\tilde D_{\nu_1}^2} {\tilde D_{\nu_2}^2-\tilde D_{\nu_1}^2},\\
&({\rm IH}): \ \Delta m^2_{\rm atm}\equiv  \Delta m^2_{\rm sol}   \frac{\tilde D_{\nu_2}^2-\tilde D_{\nu_3}^2}{\tilde D_{\nu_2}^2-\tilde D_{\nu_1}^2}.
\end{align}

\noindent The sum of neutrino masses is constrained by the minimal standard cosmological model $\Lambda_{\rm CDM} + \sum D_{\nu}$ that provides the upper bound $\sum D_{\nu}\le$ 120 meV~\cite{Vagnozzi:2017ovm, Planck:2018vyg}, although it becomes weaker if the data are analyzed in the context of extended cosmological models~\cite{ParticleDataGroup:2014cgo}.
%%%
Recently, DESI and CMB data combination provides more stringent upper bound on the sum as $\sum D_{\nu}\le$ 72 meV~\cite{DESI:2024mwx}. 
%%%
Direct search for neutrino mass is done by the Karlsruhe Tritium Neutrino (KATRIN) experiment~\cite{KATRIN:2021uub},
which is the first sub-eV sensitivity on $m_{\nu_e}^2=(0.26\pm0.34)$ eV$^2$ at 90 \% CL.
Here,
$m_{\nu_e}^2$%\equiv \sum_i|U_{ei}|^2 D^2_{\nu_i}$
is defined by
\begin{align}
\kappa^2_\nu 
\left[(\tilde D_{\nu_1} \cos\zeta_{13}\cos\zeta_{12})^2+(\tilde D_{\nu_2} \cos\zeta_{13}\sin\zeta_{12})^2+(\tilde D_{\nu_3} \sin\zeta_{13})^2\right].
\end{align}
 $\zeta_{12,23,13}$ are mixing angles for $U_{\rm PMNS}$ in the standard parametrization.
 Each of mixing angles are obtained by
\begin{align}
\sin\zeta_{13} = |(U_{\rm PMNS})_{13}|,\quad \tan\zeta_{23}\equiv \frac{(U_{\rm PMNS})_{23}}{(U_{\rm PMNS})_{33}},\quad
 \tan\zeta_{12}\equiv \frac{(U_{\rm PMNS})_{12}}{(U_{\rm PMNS})_{11}}.
 \end{align}
 
%We also take this experiment in account in our numerical analysis below.
\noindent In numerical analysis, we will adopt NuFit 6.1~\cite{Esteban:2024eli} to experimental ranges of neutrino observables in addition to the above constraints of neutrino masses.

\section{Phenomenology}
\subsection{Lepton Flavor Violations and lepton $g-2$}
\noindent Branching ratio of $\ell_i \to \ell_j \gamma$ and lepton $g-2$ are arisen from the same coupling of $f_{i\alpha}$ at the one-loop level, and their forms are given by
\begin{align}
&{\rm BR} (\ell_i \to \ell_j \gamma) \approx \frac{48 \pi^3 \alpha_{\rm em}}{G_F^2 m^2_{\ell_i}} C_{ij} \left( \left| a_{L_{ij}} \right|^2 + \left| a_{R_{ij}} \right|^2 \right) \, , \label{eq:lfvs} \\
&\Delta a_{\ell_i} \approx - \frac{m_{\ell_i}}{2} \left(a_{L_{ii}} + a_{R_{ii}} \right) \, , \label{eq:da}
\end{align}
where $\ell_1 \equiv e$, $\ell_2 \equiv \mu$, $\ell_3 \equiv \tau$, $\alpha_{\rm em} \approx 1/137$ is the fine structure constant, and $G_F \approx 1.17 \times 10^{-5} \, {\rm GeV}^{-2}$ is the Fermi constant. 
{$C_{ij}$ is defined by $C_{21} \approx 1$, $C_{31} \approx 0.1784$, $C_{32} \approx 0.1736$, otherwise $C_{ab}$ is zero. Here $a_{L,R}$ are given by
\begin{align}
a_{L_{ij}} &= -m_{\ell_i}  \sum_{\alpha = 1}^3 \frac{f_{i\alpha} f^\dag_{\alpha j}}{(4 \pi)^2}  \int[dx]_3 
\frac{xz}{(x + y) m_\eta^2 + z M^2_{\psi_\alpha} - xz m_{\ell_i}^2 - yz m_{\ell_j}^2 }\\
&\approx
 - m_{\ell_i}  \sum_{\alpha = 1}^3 \frac{\tilde f_{i\alpha}\tilde  f^\dag_{\alpha j}}{(4 \pi)^2} \frac{|f_{33}|^2}{M^2_{\psi_3}} 
 \int[dx]_3 
\frac{xz}{(x + y) \tilde m_\eta^2 + z \tilde M^2_{\psi_\alpha} }, 
 \\
a_{R_{ij}} &= 
-m_{\ell_j}  \sum_{\alpha = 1}^3 \frac{f_{i\alpha} f^\dag_{\alpha j}}{(4 \pi)^2}  \int[dx]_3 
\frac{yz}{(x + y) m_\eta^2 + z M^2_{\psi_\alpha} - xz m_{\ell_i}^2 - yz m_{\ell_j}^2 }\\
&\approx
-m_{\ell_j}  \sum_{\alpha = 1}^3 \frac{\tilde f_{i\alpha} \tilde f^\dag_{\alpha j}}{(4 \pi)^2}  \frac{|f_{33}|^2}{M^2_{\psi_3}}  \int[dx]_3 
\frac{yz}{(x + y) \tilde m_\eta^2 + z \tilde M^2_{\psi_\alpha} }, 
\end{align}
where  $[dx]_3 \equiv dx dy dz \delta (1 - x - y - z)$.
% and $m_{\ell_i}, m_{\ell_j} \ll m_\eta, M_\psi$ is assumed in the last equations. 
 % 
The current upper bounds on ${\rm BR} (\ell_a \to \ell_b \gamma)$ are given at 90\% CL. as~\cite{MEGII:2025gzr, MEGII:2023ltw, BaBar:2009hkt, Belle:2021ysv}
\begin{align}
{\rm BR} (\mu \to e \gamma) < 1.5 \times 10^{-13} \, , \quad {\rm BR} (\tau \to e \gamma) < 3.3 \times 10^{-8} \, , \quad {\rm BR} (\tau \to \mu \gamma) < 4.2 \times 10^{-8} \, .
\end{align}
%Among these bounds, ${\rm BR} (\mu \to e \gamma)$ is relevant to our discussion. 
The lepton $g-2$ receives the following experimental measurements~\cite{Fan:2022oyb,Fan:2022eto,ParticleDataGroup:2024cfk,Morel:2020dww}:
\begin{align}
\Delta a_e = (3.41 \pm 1.64) \times 10^{-13} \, , \quad \Delta a_{\mu} = (39 \pm 64) \times 10^{-11} \, .
\label{eq:g-2ell}
\end{align}
Here, $\Delta a_{\tau}$ is not relevant since it is less precise compared with $\Delta a_{e, \mu}$.
Therefore, we neglect analysis of $\Delta a_{\tau}$ in the numerical analysis.

\subsection{Dark Matter Analysis}
\noindent In this scenario, the model admits two viable DM candidates: bosonic state $S$ and a fermionic state $\psi_1$.
We notice that $\eta_0$ cannot be a DM candidate since it interacts with the Z-boson, which induces sizable spin-independent elastic scattering cross sections off nuclei. The corresponding scattering rates exceed the stringent bounds imposed by current direct detection experiments, thereby excluding $\eta_0$ as a phenomenologically acceptable DM candidate.

\subsubsection{Bosonic DM case: $S$ state}
\noindent Since $S$ does not couple to Standard Model fields through Yukawa interactions, its only portal to the visible sector arises $via$ scalar interactions in the Higgs potential. In such a case, detailed analysis is already explored, and the correct relic density of  DM is obtained near the half of Higgs mass by the resonant enhancement~\cite{Kanemura:2010sh} without conflict with the direct detection bounds. Therefore, 
\begin{align} 
m_S \approx \frac{m_h}{2} \approx 63 \ {\rm GeV}.
\end{align}
In our numerical analysis, we adopt this fixed mass for $m_S< M_{\psi_1}$.

Here, it must be noted that $S$ can decay into $H$ and $\eta$ through one-loop level, when this process is kinematically allowed.
The corresponding terms are $\kappa$ and $\lambda_{0}$ and the decay rate $(\Gamma_s)$ is found as
\begin{align}
& \Gamma_s =\frac{|\mu_{SH\eta}|^2}{16\pi m_S^2} \sqrt{1-4\frac{m_h^2}{m^2_S}},\\
& \mu_{SH\eta} = -\frac{\kappa \lambda_{0}^*}{16\pi^2} 
\left[
2+ \frac{1+r}{1-r}\ ln (r)
\right],
\end{align}
where $r\equiv m^2_S/\Lambda^2$, $\Lambda$ being cut-off scale, putting $m_\eta\sim m_h$ for simplicity.
In order to satisfy that DM has to be longer the age of Universe ${\cal O}(10^{41})\ {\rm GeV}^{-1}\lesssim\Gamma_s^{-1}$, the following condition under $m_S=(m_h-1)/2$ has to be satisfied.
\begin{align}
\Lambda&=10^5\ {\rm GeV},\quad \kappa \lambda_{0}^*\lesssim 6.47\times 10^{-18} \ {\rm GeV},\\
\Lambda&=10^{10}\ {\rm GeV},\quad \kappa \lambda_{0}^*\lesssim 2.31\times 10^{-18} \ {\rm GeV},\\
\Lambda&=10^{19}\ {\rm GeV},\quad \kappa \lambda_{0}^*\lesssim 1.07\times 10^{-18} \ {\rm GeV}.
\end{align}
%
%
%
\if0
Here, it must be noted that $S$ can decay into two Higgs pairs through one-loop level. The corresponding terms are $\kappa$ and $\lambda_{HS}$ and the decay rate $(\Gamma_s)$ is found as
\begin{align}
& \Gamma_s =\frac{|\mu_{SHH}|^2}{16\pi m_S^2} \sqrt{1-4\frac{m_h^2}{m^2_S}},\\
%
& \mu_{SHH} = -\frac{\kappa \lambda_{HS}}{16\pi^2} 
\left[
2+ \frac{1+r}{1-r}\ ln (r)
\right],
\end{align}
where $r\equiv m^2_S/\Lambda^2$, $\Lambda$ being cut-off scale, putting $m_\eta\sim m_h$ for simplicity.
In order to satisfy that DM has to be longer the age of Universe ${\cal O}(10^{41})\ {\rm GeV}^{-1}\lesssim\Gamma_s^{-1}$, the following condition under $m_S=(m_h-1)/2$ has to be satisfied.
\begin{align}
\Lambda&=10^5\ {\rm GeV},\quad \kappa \lambda_{HS}\lesssim 6.47\times 10^{-18} \ {\rm GeV},\\
%
\Lambda&=10^{10}\ {\rm GeV},\quad \kappa \lambda_{HS}\lesssim 2.31\times 10^{-18} \ {\rm GeV},\\
%
\Lambda&=10^{19}\ {\rm GeV},\quad \kappa \lambda_{HS}\lesssim 1.07\times 10^{-18} \ {\rm GeV}.
\end{align}
\fi

\subsubsection{Fermionic DM case: $\psi_1$ state}
%
%\noindent \textcolor{blue}{The lightest exotic fermion $\psi_1$ is stable on cosmological timescales since no renormalizable operator inducing its decay is generated in our framework.} 
{
Although the fermionic DM  can decay into two Higgs pair plus lepton through the Yukawa coupling $f$ and the same diagram as the $D$ decay process,
we suppose that the fermionic dark matter candidate is effectively stable on cosmological timescales since its lifetime exceeds the age of the Universe
once the S decay process is suppressed enough as discussed in the bosonic DM section.
}
Then, the mass of $X$ is given by $M_{\psi_1}$. Since $X$ does not directly couple to quark sector, we can neglect the constraints of direct detection searches. 
The relevant annihilation processes are $X\bar X \to \ell_i \bar{\ell_j} (\nu_\sigma\bar\nu_\rho)$ and they are s-wave dominant due to the Dirac fermion DM.
The thermally averaged cross section in the limit of massless final states is given by
\begin{align}
&    \langle \sigma v \rangle(X\bar X \to \nu_\sigma\bar\nu_\rho) \approx \frac{|f_{33}|^4}{M^2_{\psi_3}} \frac{ \tilde M_{\psi_1}^2}{32\pi} 
    \frac{c^2_\theta(\tilde m_1^2 + \tilde M_{\psi_1}^2)+s^2_\theta(\tilde m_2^2 + \tilde M_{\psi_1}^2)} 
    {(\tilde m_1^2 + \tilde M_{\psi_1}^2)^2(\tilde m_2^2 + \tilde M_{\psi_1}^2)^2} 
    \sum_{\sigma,\rho=1}^3 |\tilde f'_{\sigma 1} \tilde f'^\dag_{1\rho}|^2,
    \\
&    \langle \sigma v \rangle(X\bar X \to  \ell_i \bar{\ell_j}) 
    \approx \frac{|f_{33}|^4}{M^2_{\psi_3}} \frac{ \tilde M_{\psi_1}^2}{32\pi (\tilde m_\eta^2 + \tilde M_{\psi_1}^2)^2} 
%    \frac{1} {(\tilde m_\eta^2 + \tilde M_{\psi_1}^2)^2} 
    \sum_{\sigma,\rho=1}^3 |\tilde f_{i 1} \tilde f^\dag_{1j}|^2,   
\end{align}
where $f'_{a1}\equiv \sum_{i=1}^3 (U^\dag_{\rm PMNS})_{ai} f_{i1}$.
The value of cross section  to satisfy the observed relic density of DM is approximately given by
\begin{align}
1.776\lesssim \frac{  \langle \sigma v \rangle \times 10^9}{{\rm GeV}^{-2}}\lesssim 1.970,
\end{align}
at 2$\sigma$ confidence level (C.L.) that corresponds to $\Omega h^2=0.1196\pm2\times 0.0031$~\cite{Planck:2013pxb}.
Spin independent scattering cross section can be found at one-loop level via gauge bosons such as photon and Z boson.
However, it can be negligible when the Yukawa couplings are of the order $10^{-3}$.

\section{Numerical analysis}
%Before performing our numerical analysis, we fix the method.
\noindent In this section, we perform a $\Delta \chi^2$ numerical analysis.
Here, we assume Gaussian Distribution with four observables; $[s_{{12}},s_{{13}},\Delta m^2_{\rm sol},\Delta m^2_{\rm atm}]$ from NuFit 6.1~\cite{Esteban:2024eli}, as follows:
\begin{equation}
\Delta \chi^2 = \sum_{i}  \left( \frac{O_i^{\rm obs} - O_i^{\rm th}}{\delta O_i^{\rm exp}} \right)^2, \label{eq:chi-square}
\end{equation}
where $O_i^{\rm obs (th)}$ is observed (theoretically obtained) value of corresponding observables and $\delta O_i^{\rm exp}$ expresses experimental error.
Four degrees of freedom lead us to  
$\Delta \chi^2=4.72$ at 1$\sigma$, $\Delta \chi^2=9.72$ at 2$\sigma$, $\Delta \chi^2=16.25$ at 3$\sigma$, $\Delta \chi^2=24.50$ at 4$\sigma$, and $\Delta \chi^2=34.56$ at 5$\sigma$, respectively.
% following C.L.; 
%
%
It is notable that we do not consider  Dirac phase denoted by $\delta_{\rm CP}$ and $s_{{23}}$ as output parameters.
Because any values of $\delta_{\rm CP}$ are allowed by experiments at 3$\sigma$ C.L..
While $s_{23}$ is deviated from the Gaussian distribution and we independently impose to be the following experimental constraint
\begin{align}
&0.432\le s^2_{23}\le0.587\ {\rm for \ NH}, \\
& 0.437\le s^2_{23}\le0.590\ {\rm for \ IH},
\end{align}
within the range of 3$\sigma$ C.L. in Nufit 6.1.
In a numerical analysis, we plot the allowed parameter space with green(red) points within $5(3)\sigma$ C.L.. 
In addition to the allowed regions in Nufit 6.1, we take into account the recent experimental result from JUNO~\cite{JUNO:2025gmd} at the 1$\sigma$ C.L.:
\begin{align}
\sin^2\theta_{12} =0.3092\pm{0.0087},\quad \frac{\Delta m^2_{\rm sol}}{10^{-5}{\rm eV}^2} =7.50\pm 0.12.
\end{align}
In our numerical analysis below, blue circles are shown when both experimental constraints are satisfied.

\noindent We randomly select our viable input parameters  as follows:
\begin{align}
& |\tilde f_{11,12,13,21,22,23,31,32}, \ \tilde g_{11,12,13,21,22,23,31,32}, g_{33}| =[10^{-3},\sqrt{4\pi}], \quad s_\theta=[-1,1],\\
%%%
& M_{\psi_3} =[10,10^5]\ {\rm GeV},\quad [\tilde M_{\psi_1},  \tilde M_{\psi_2}, \tilde m_1 ,  \tilde m_2]=[10^{-5},0], 
\end{align}
where we set $m_\eta=m_2$ simply to evade the constraint of oblique parameters.

\subsection{NH}
\subsubsection{Bosonic DM: $S$ state}

\noindent In case of bosonic DM, we impose the following condition 
\begin{align}
\tilde m_1 \le 1.2 \times (\tilde M_{\psi_1}, \tilde m_2).
\end{align}
where $1.2$ is roughly evaluated by suppressing co-annihilation cross sections~\cite{Griest:1990kh}.
%

%%%%%%%%%%%%%%%%%%%
\begin{figure}[tb]\begin{center}
\includegraphics[width=77mm]{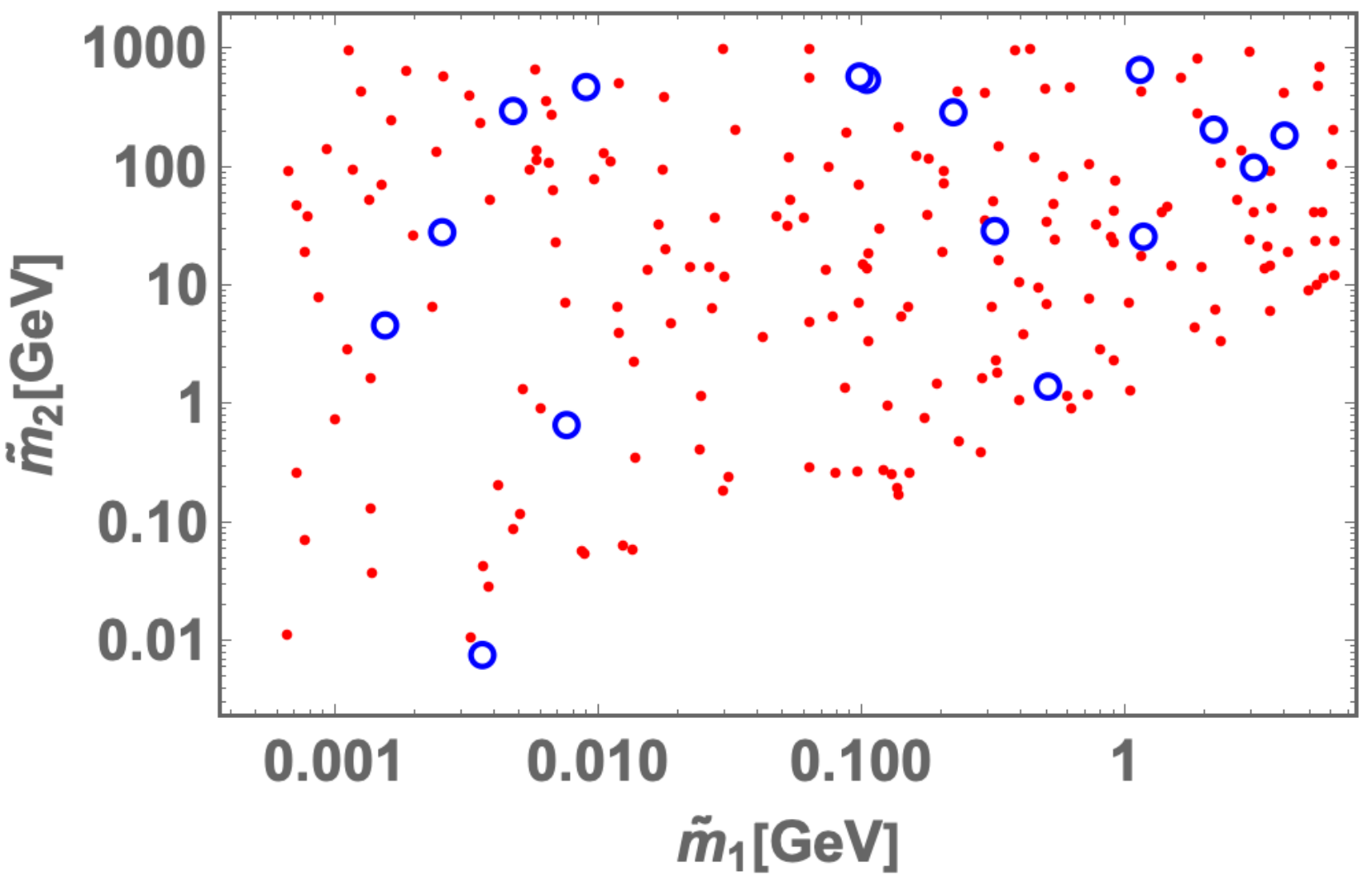}
\includegraphics[width=77mm]{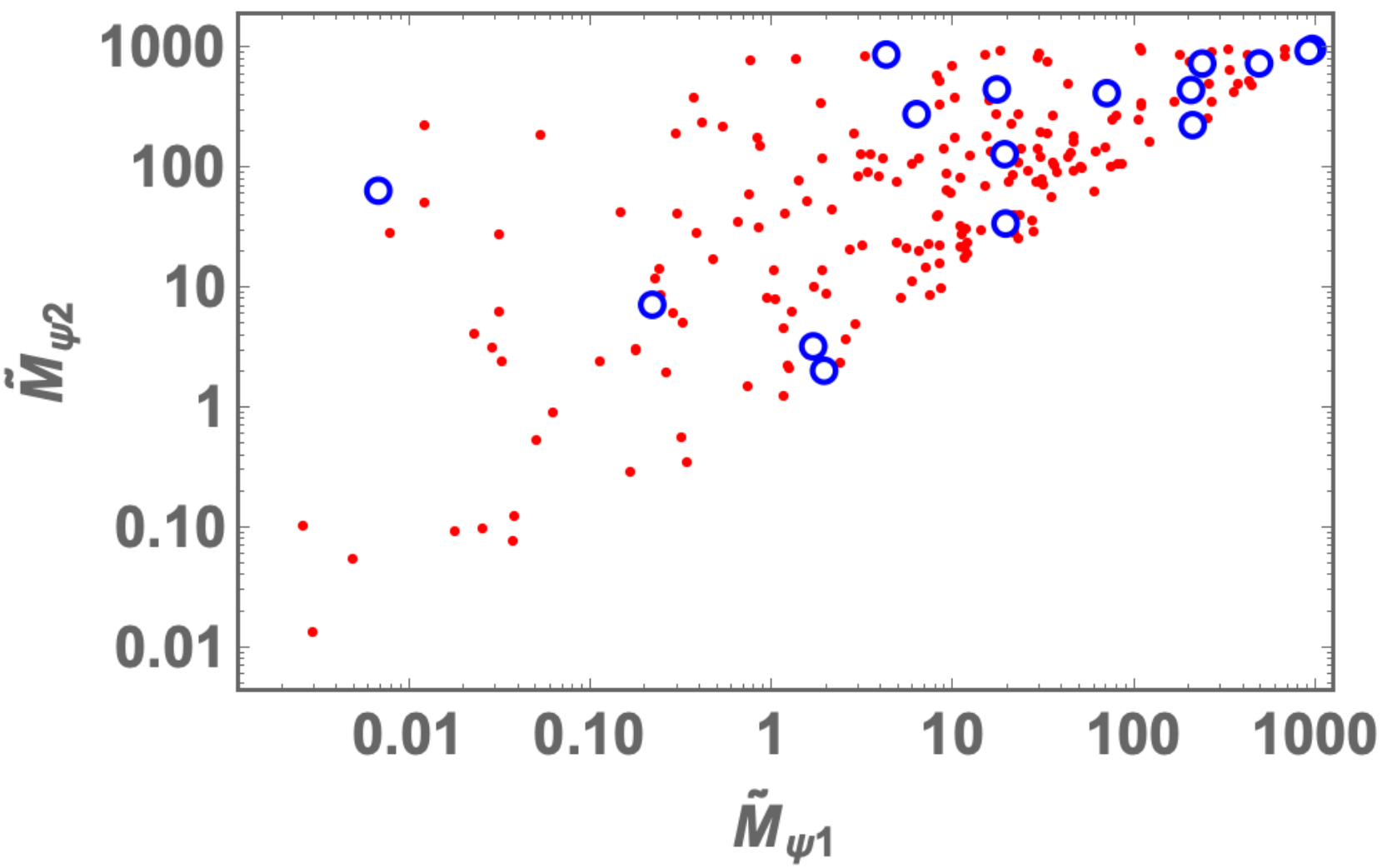}\\
\includegraphics[width=77mm]{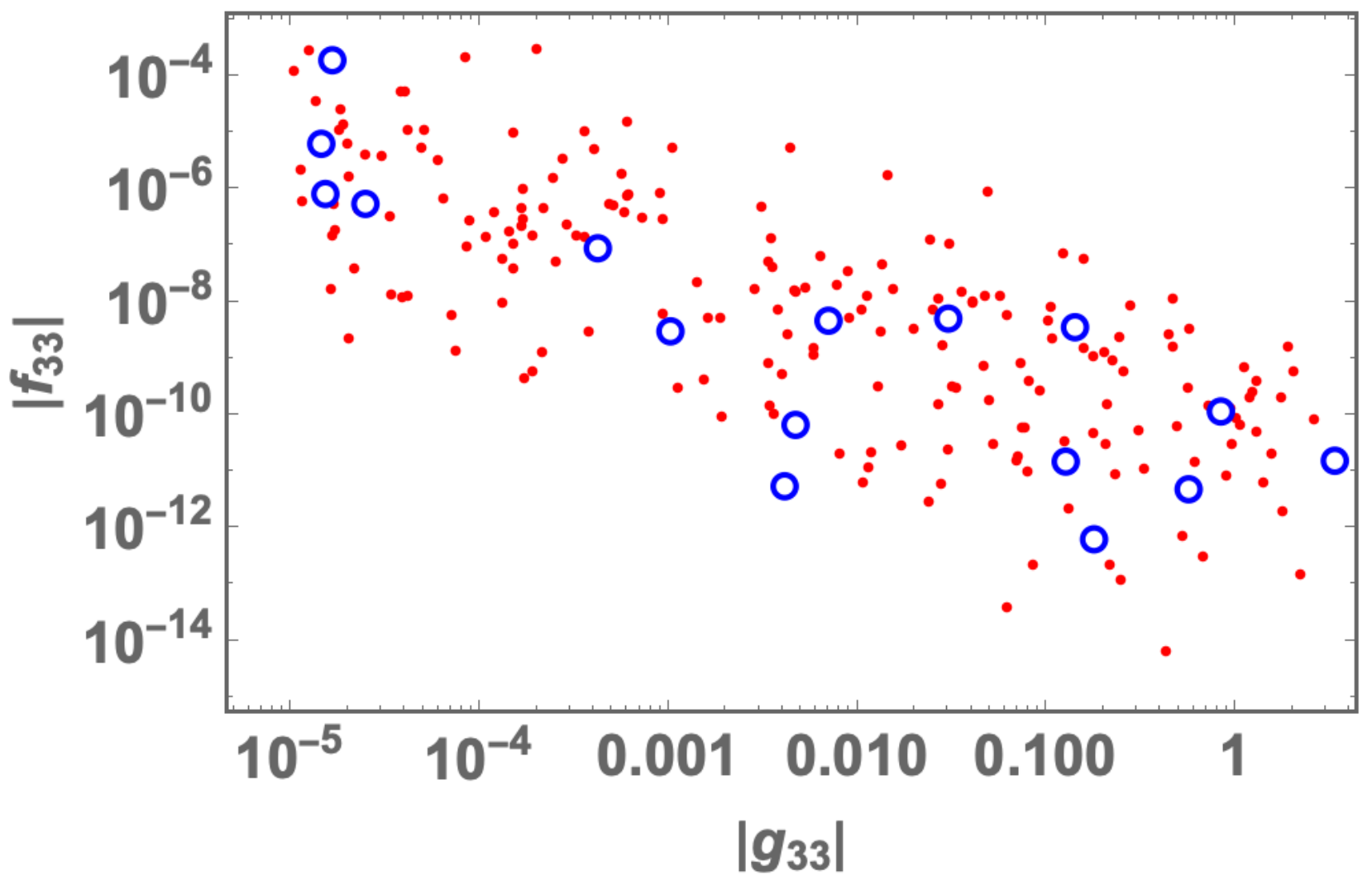}
\includegraphics[width=77mm]{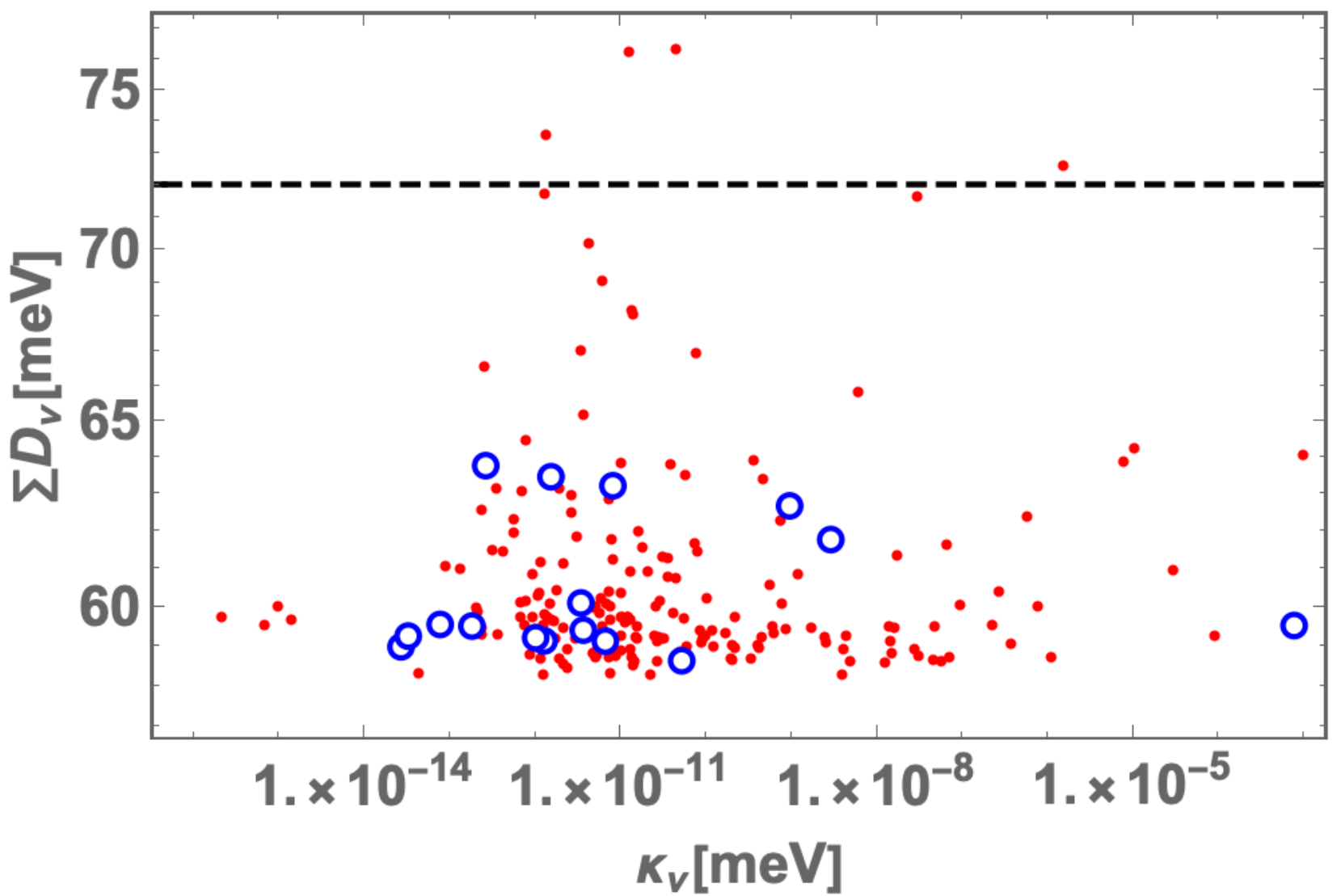}
\caption{The allowed regions of $\tilde m_{1,2}$ (left-up panel), $\tilde M_{\psi_{1,2}}$ (right-up panel), 
 $| f_{33}|-|g_{33}|$ (left-down panel),  and  $\sum D_{\nu}-|\kappa_\nu|$ in meV unit (right-down panel), respectively.
The dashed horizontal line in the right-down plot represents the upper bound on $\sum D_\nu\le72$ meV.
As we explained in the first part of this section, the red color is within $5\sigma$ C.L..
%, while the green one is within $3\sigma$ C.L..
The blue circle satisfies the recent experimental result of JUNO in addition to the Nufit 6.1. }
\label{fig:nhbdm1}
\end{center}\end{figure}
%%%%%%%%%%%%%%%%%%%
%
\noindent Fig.~\ref{fig:nhbdm1}, we show the allowed regions of $\tilde m_{1,2}$ (left-up panel), $\tilde M_{\psi_{1,2}}$ (right-up panel), 
 $| f_{33}|-|g_{33}|$ (left-down panel),  and  $\sum D_{\nu}-|\kappa_\nu|$ in meV unit (right-down panel), respectively.
The dashed horizontal line in the right-down plot represents the upper bound on $\sum D_\nu\le72$ meV.
{
%As we explained in the first part of this section, 
The red color is within $5\sigma$ C.L., which corresponds to $\Delta\chi^2\approx34.56$,
%, while the green one is within $3\sigma$ C.L..
while the blue circle satisfies the recent experimental result of JUNO in addition to the Nufit 6.1 within $5\sigma$ C.L..}
One straightforwardly finds that the allowed parameter space is drastically reduced one the JUNO results are taken into account.
JUNO results suggest that $\sum D_\nu$ is localized at nearby 59$-$64 meV, which is totally safe for the experimental upper bounds from DESI.

%%%%%%%%%%%%%%%%%%%
\begin{figure}[tb]\begin{center}
\includegraphics[width=77mm]{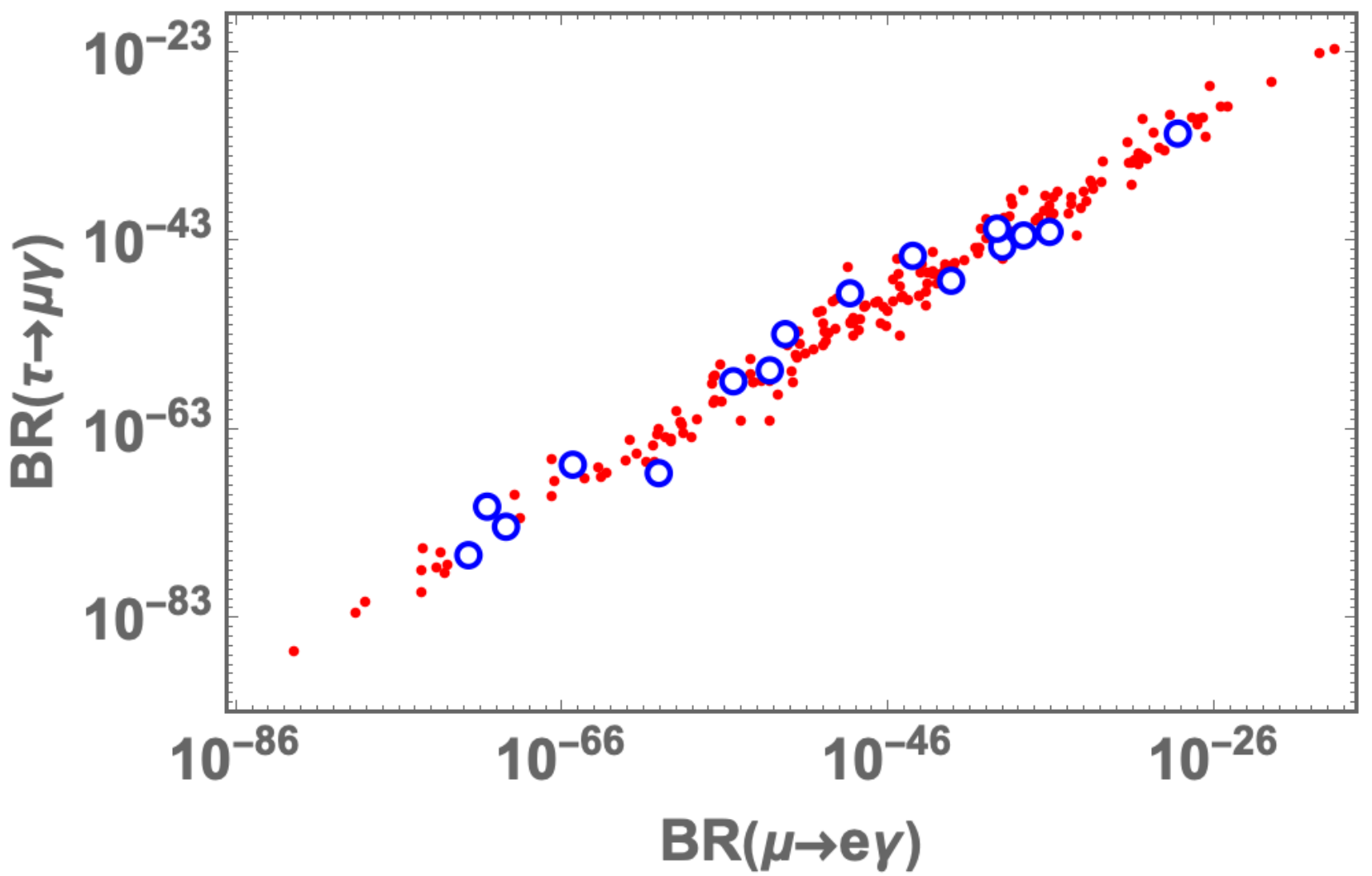}
\includegraphics[width=77mm]{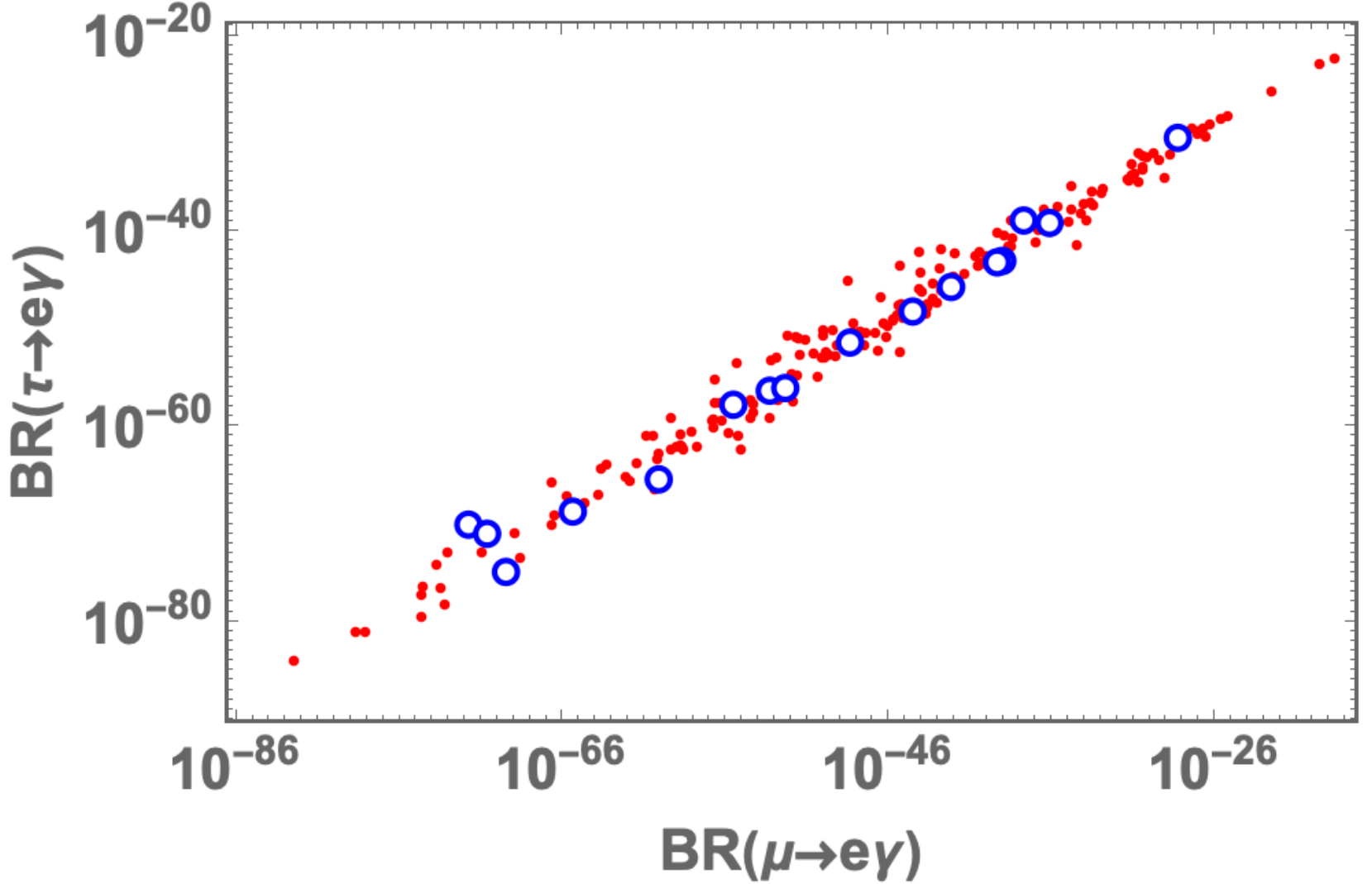}\\
\includegraphics[width=77mm]{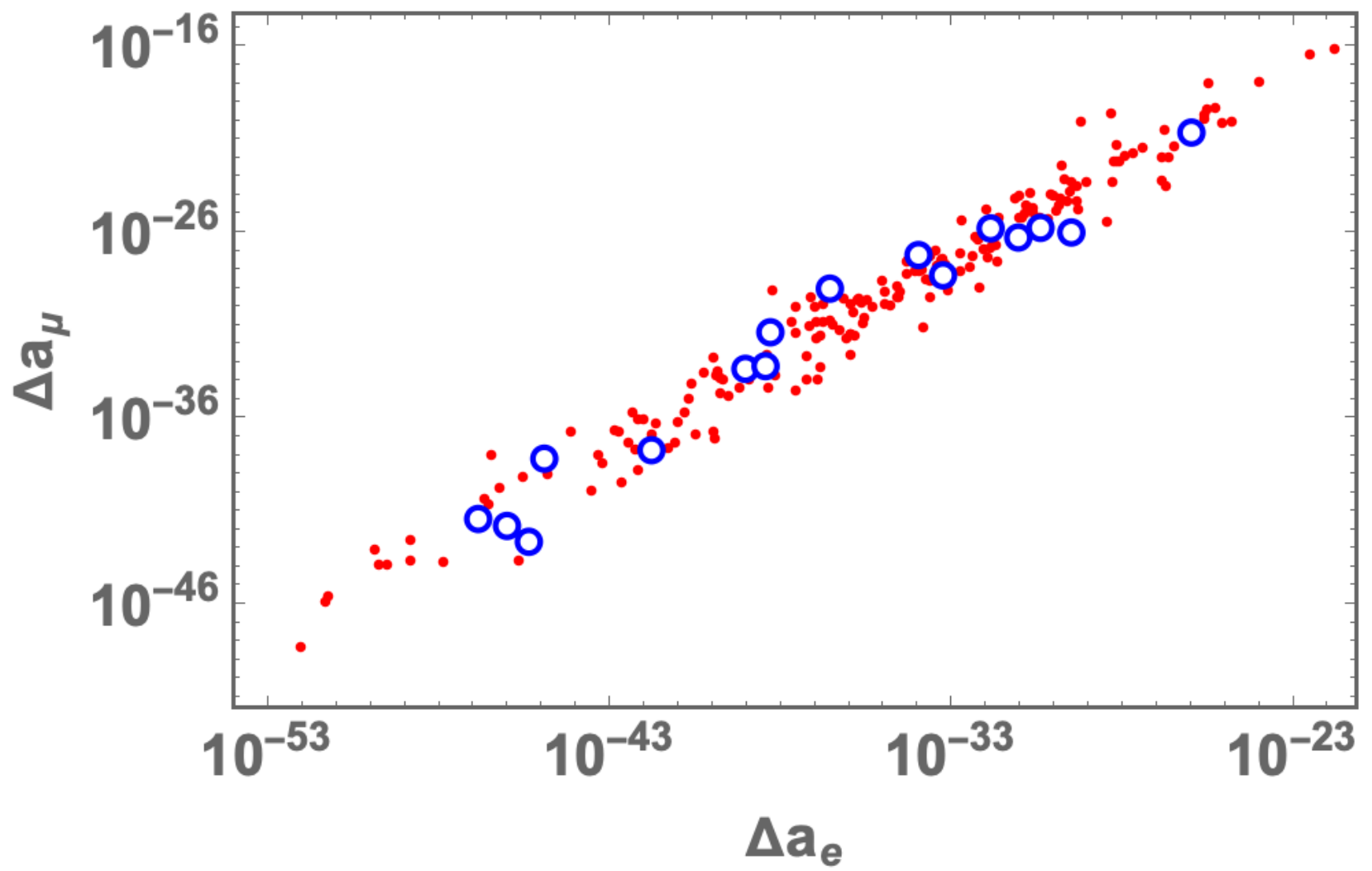}
\includegraphics[width=77mm]{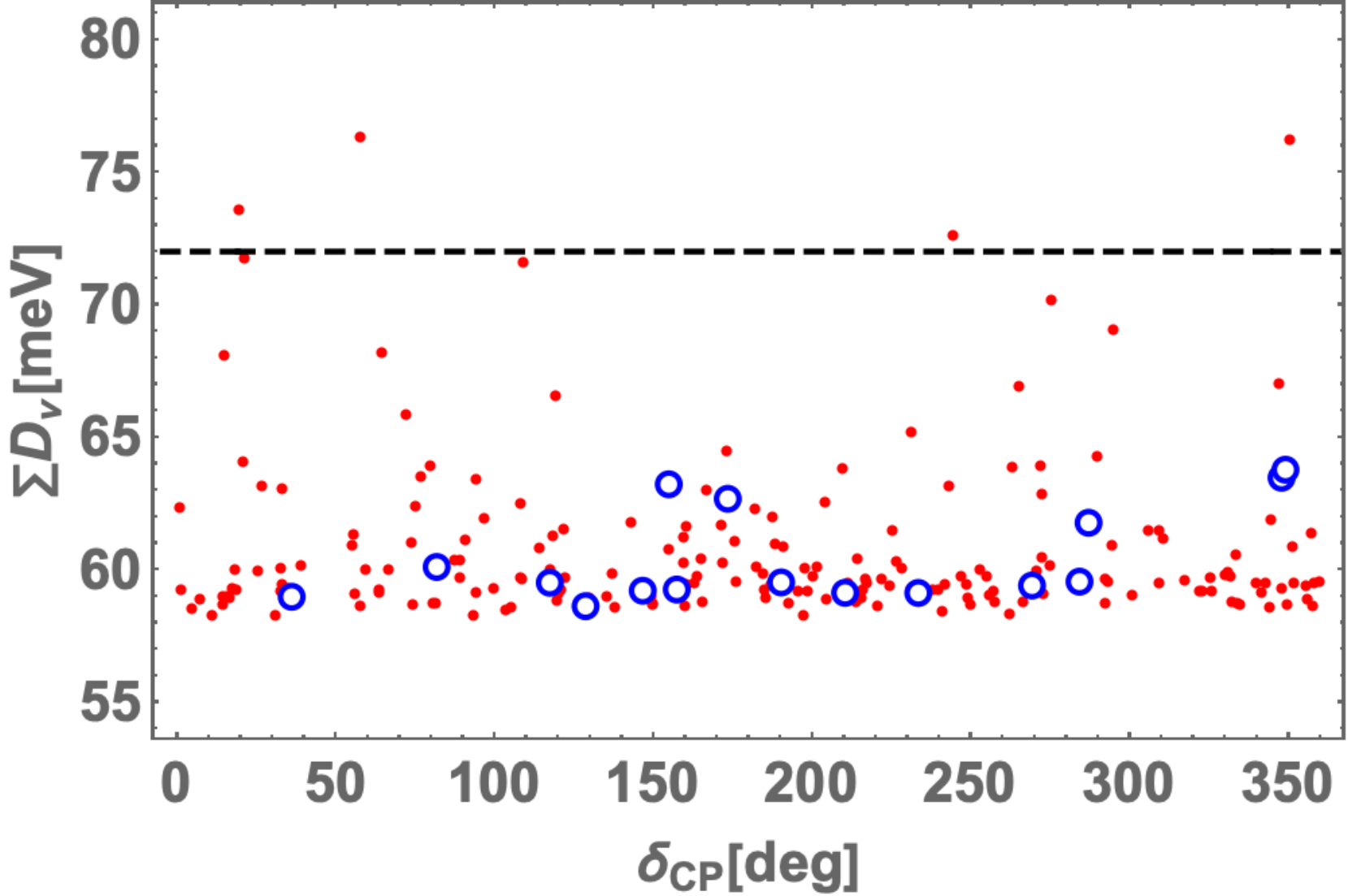}
\caption{Allowed regions of branching ratios of $\tau\to \mu\gamma$ and $\mu\to e\gamma$ (left-up panel),  $\tau\to e\gamma$ and $\mu\to e\gamma$  (right-up panel), lepton $g-2$ of 
 $\Delta a_\mu$  and $\Delta a_e$(left-down panel),  and  $\sum D_{\nu}-\delta_{\rm CP}$ in meV unit (right-down panel), respectively.
All the color legends and the line are the same as the ones in Fig.~\ref{fig:nhbdm1}.   }
\label{fig:nhbdm2}
\end{center}\end{figure}
%%%%%%%%%%%%%%%%%%%
%
\noindent In Fig.~\ref{fig:nhbdm2}, we show the allowed regions of branching ratios of $\tau\to \mu\gamma$ and $\mu\to e\gamma$ (left-up panel),  $\tau\to e\gamma$ and $\mu\to e\gamma$  (right-up panel), lepton $g-2$ of 
 $\Delta a_\mu$  and $\Delta a_e$(left-down panel),  and  $\sum D_{\nu}-\delta_{\rm CP}$ in meV unit (right-down panel), respectively.
All the color legends and the line are the same as the ones in Fig.~\ref{fig:nhbdm1}.
These figures show that the cLFVs and lepton $g-2$ are far below the experimental upper bounds.
Therefore, they are unlikely to be tested in the foreseeable future.
% testability are less verifiable.

\subsubsection{Fermionic DM: $\psi_1$ state}
\noindent In case of fermionic DM, we impose the following condition 
\begin{align}
\tilde M_{\psi_1} \le 1.2 \times (\tilde M_{\psi_2},\tilde m_1,\tilde m_2).
\end{align}
where $1.2$ is rough estimation to suppress the co-annihilation cross sections.

%%%%%%%%%%%%%%%%%%%
\begin{figure}[tb]\begin{center}
\includegraphics[width=77mm]{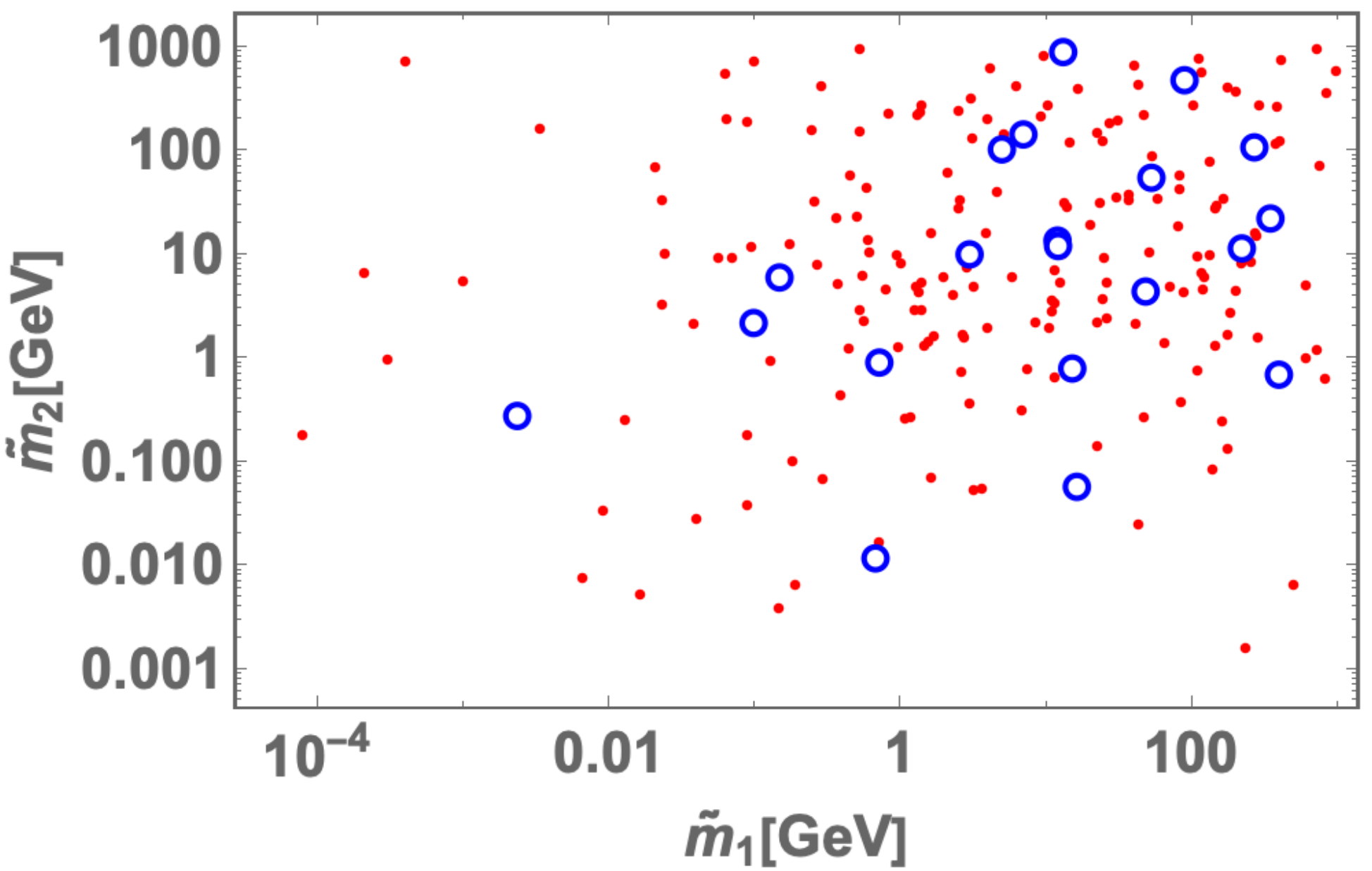}
\includegraphics[width=77mm]{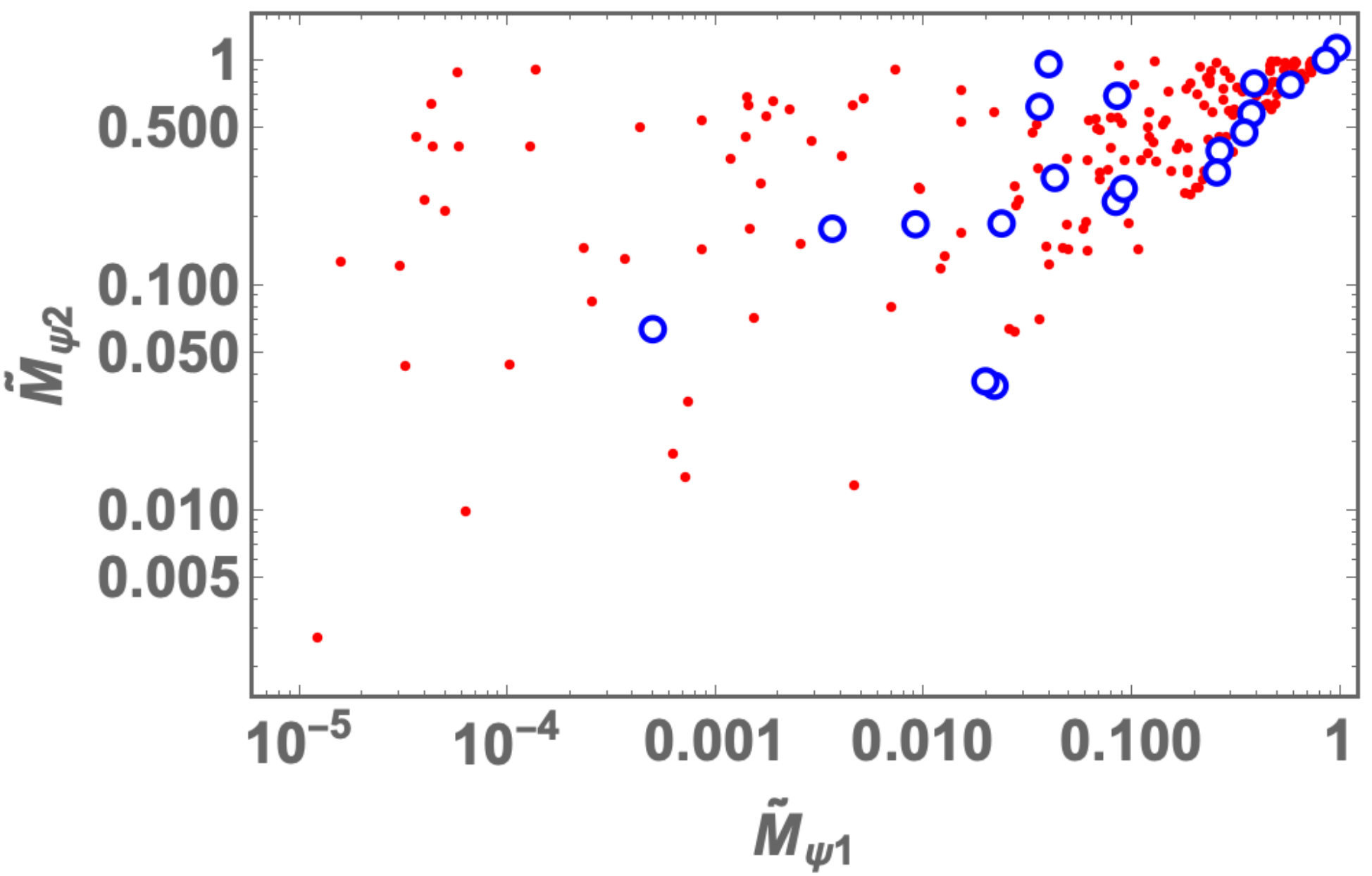}\\
\includegraphics[width=77mm]{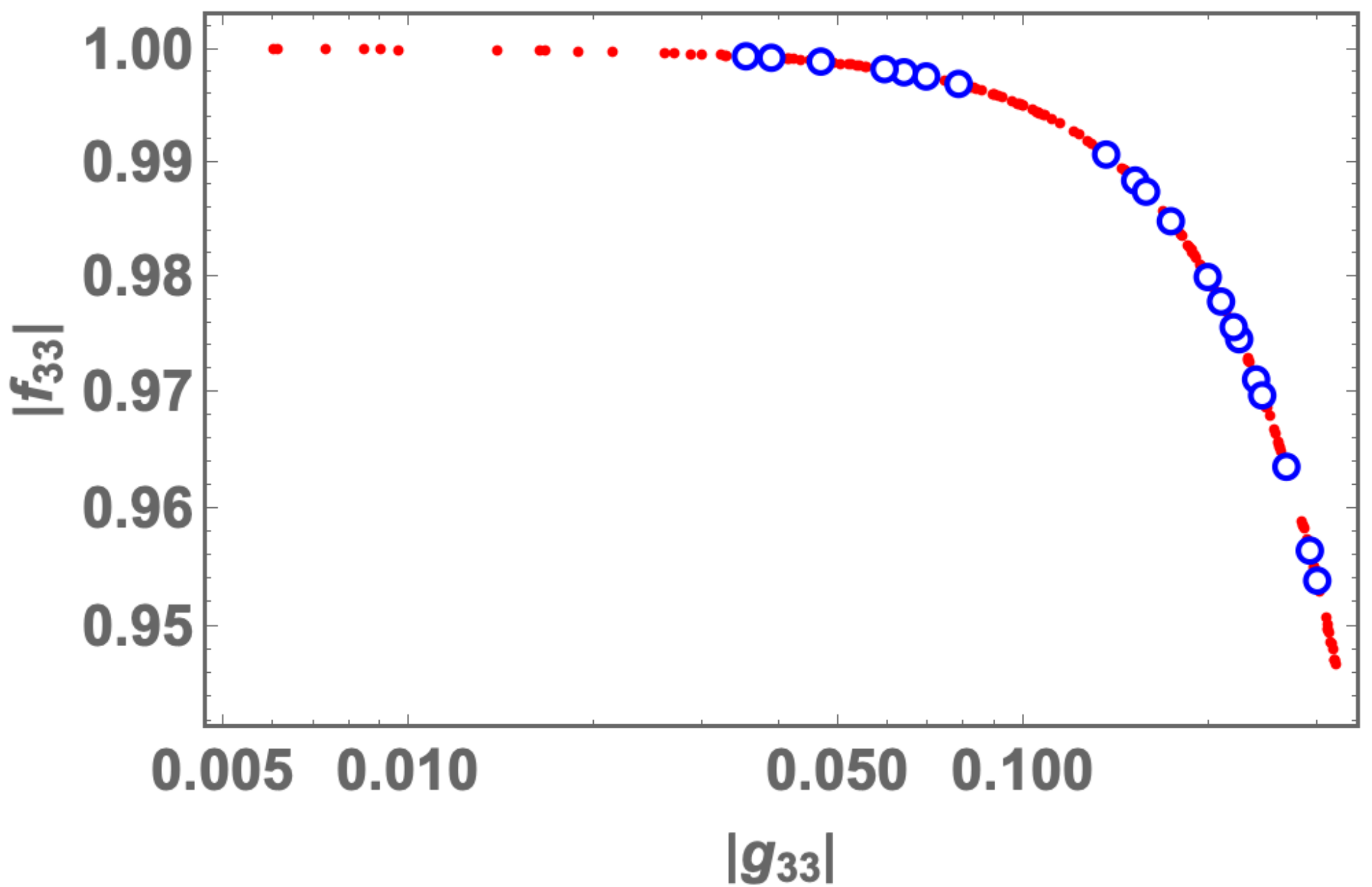}
\includegraphics[width=77mm]{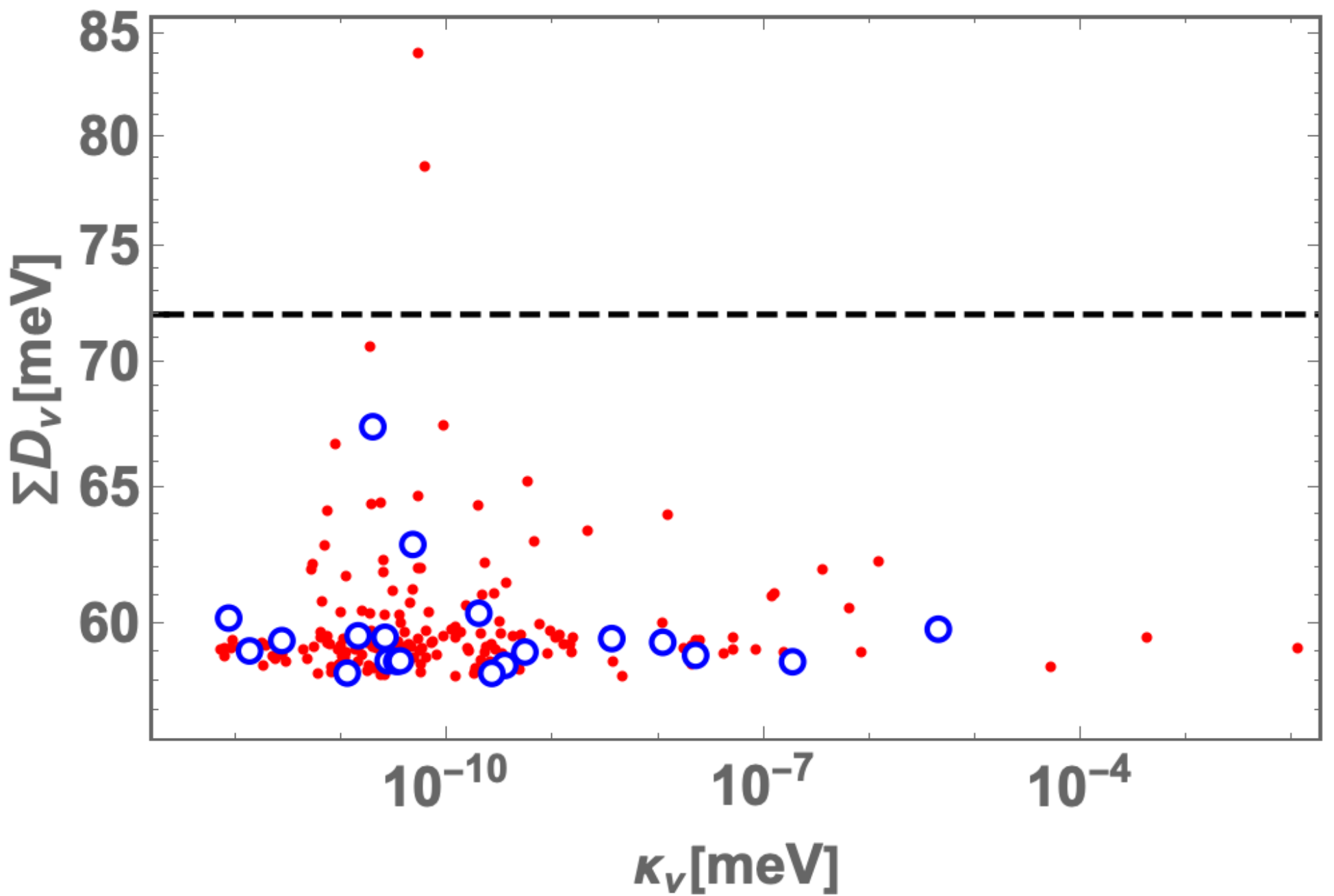}
\caption{The allowed regions of $\tilde m_{1,2}$ (left-up panel), $\tilde M_{\psi_{1,2}}$ (right-up panel), 
 $| f_{33}|-|g_{33}|$ (left-down panel),  and  $\sum D_{\nu}-|\kappa_\nu|$ in meV unit (right-down panel), respectively.
All the color legends and the line are the same as the ones in Fig.~\ref{fig:nhbdm1}.  }
\label{fig:nhfdm1}
\end{center}\end{figure}
%%%%%%%%%%%%%%%%%%%
%
\noindent Fig.~\ref{fig:nhfdm1}, we show the allowed regions of $\tilde m_{1,2}$ (left-up panel), $\tilde M_{\psi_{1,2}}$ (right-up panel), 
 $| f_{33}|-|g_{33}|$ (left-down panel),  and  $\sum D_{\nu}-|\kappa_\nu|$ in meV unit (right-down panel), respectively.
All the color legends and the line are the same as the ones in Fig.~\ref{fig:nhbdm1}.
Specifically, there is a strong correlation between $|g_{33}|$ and $|f_{33}|$, which is different from the case of BDM in NH.
JUNO results suggest that $\sum D_\nu$ is localized at nearby 59$-$68 meV, which is still safe for the experimental upper bounds from DESI.

%%%%%%%%%%%%%%%%%%%
\begin{figure}[tb]\begin{center}
\includegraphics[width=77mm]{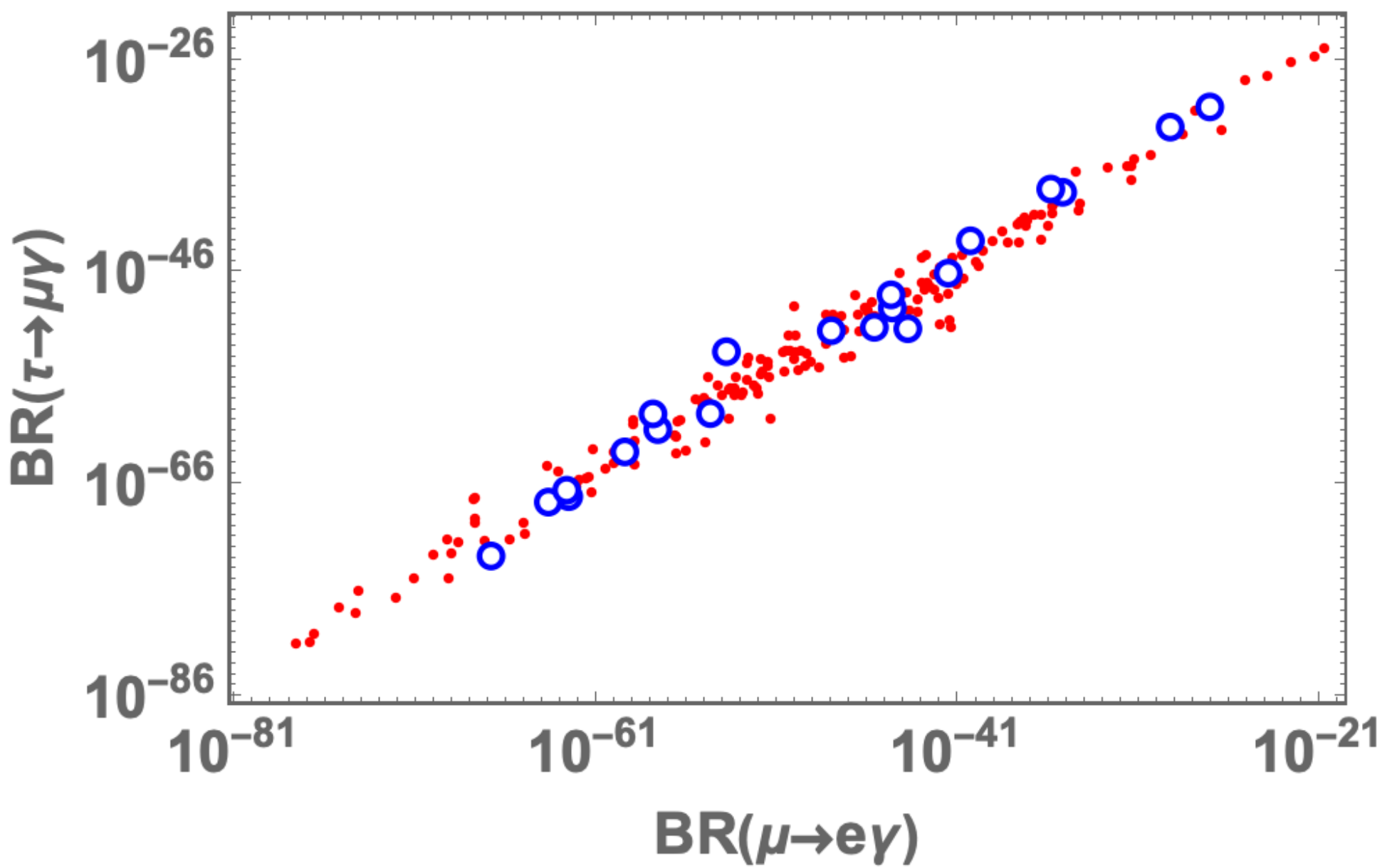}
\includegraphics[width=77mm]{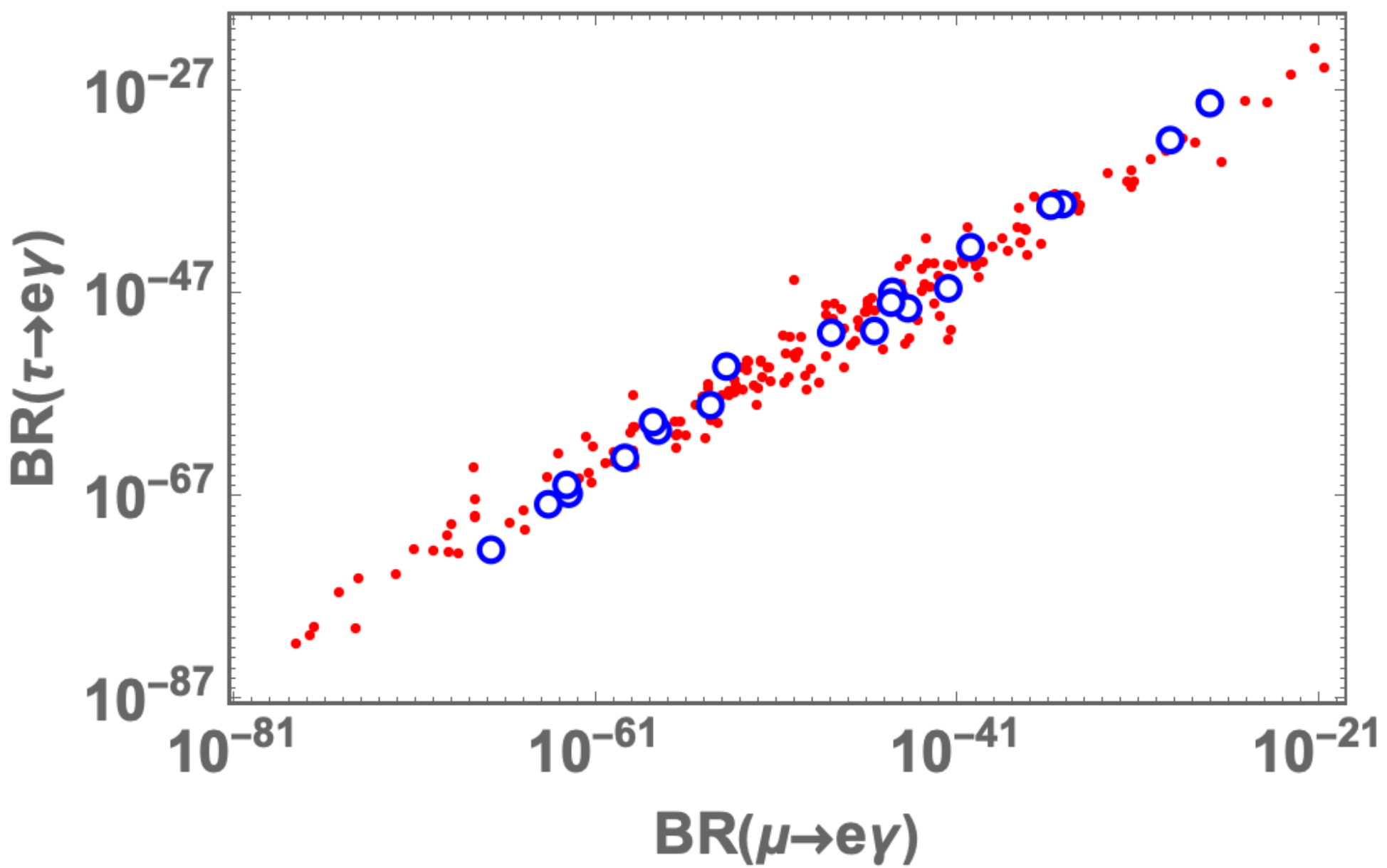}\\
\includegraphics[width=77mm]{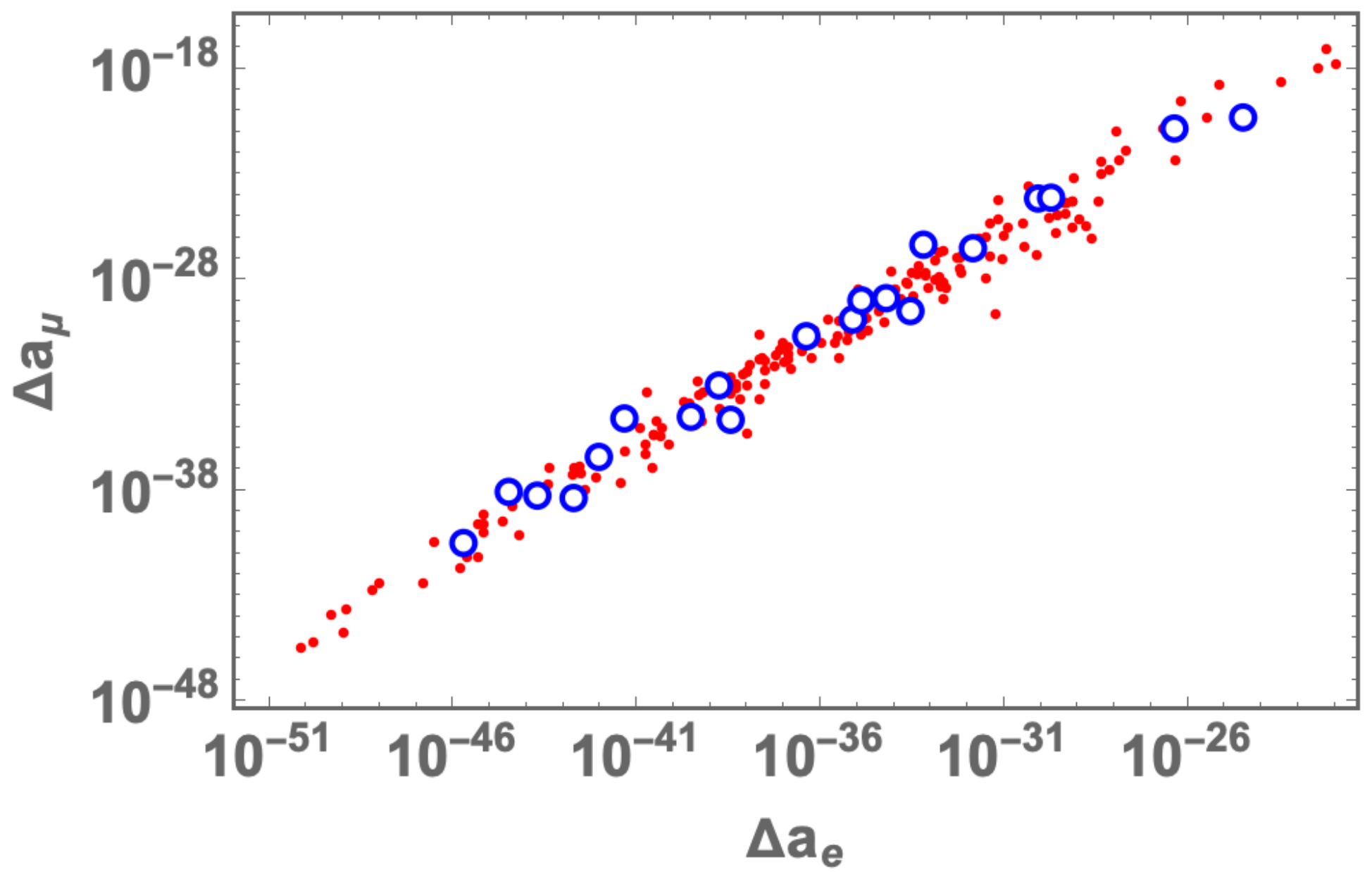}
\includegraphics[width=77mm]{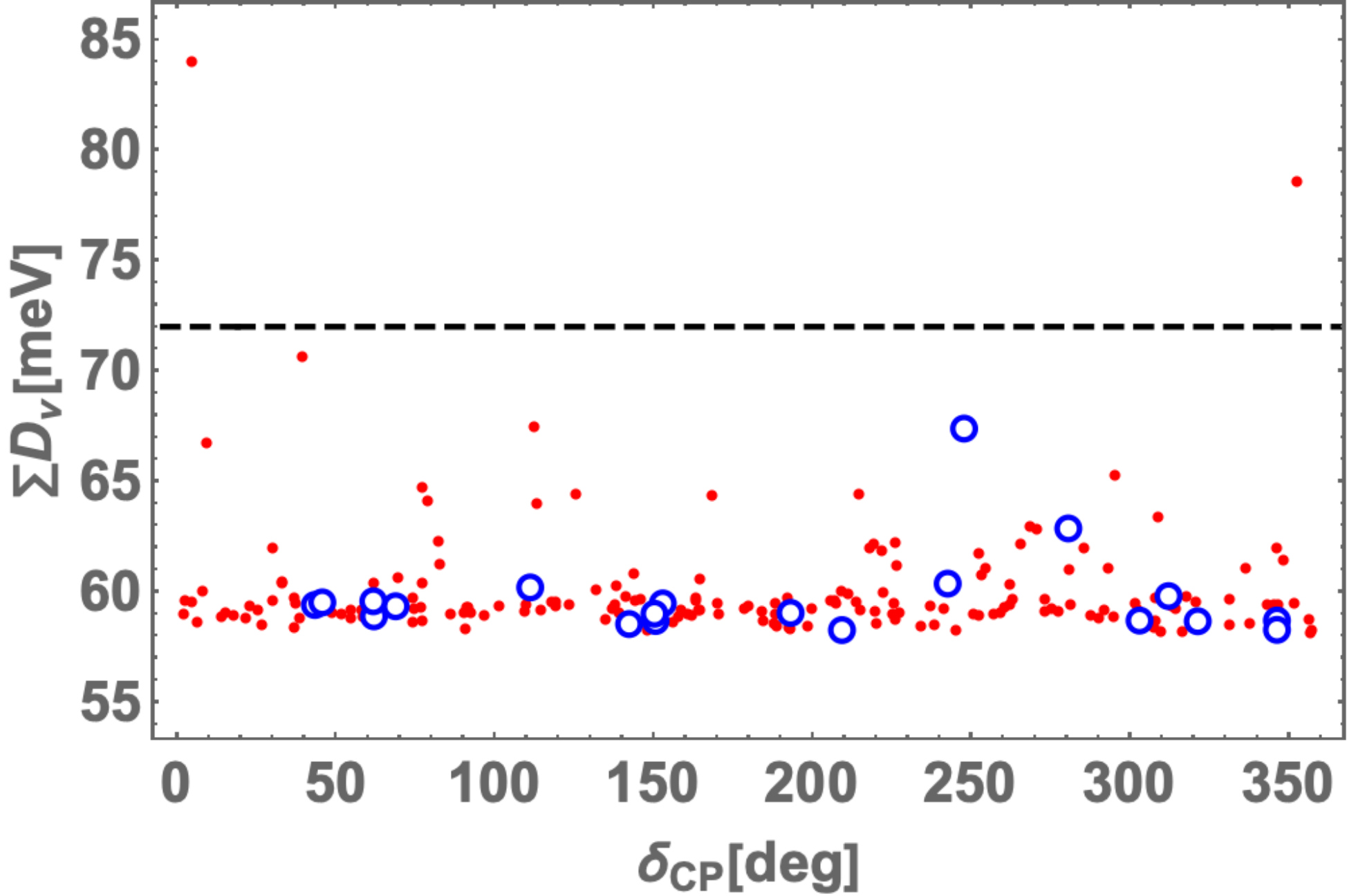}
\caption{Allowed regions of branching ratios of $\tau\to \mu\gamma$ and $\mu\to e\gamma$ (left-up panel),  $\tau\to e\gamma$ and $\mu\to e\gamma$  (right-up panel), lepton $g-2$ of 
 $\Delta a_\mu$  and $\Delta a_e$(left-down panel),  and  $\sum D_{\nu}-\delta_{\rm CP}$ in meV unit (right-down panel), respectively.
All the color legends and the line are the same as the ones in Fig.~\ref{fig:nhbdm1}.   }
\label{fig:nhfdm2}
\end{center}\end{figure}
%%%%%%%%%%%%%%%%%%%
%
\noindent In Fig.~\ref{fig:nhfdm2}, we show the allowed regions of branching ratios of $\tau\to \mu\gamma$ and $\mu\to e\gamma$ (left-up panel),  $\tau\to e\gamma$ and $\mu\to e\gamma$  (right-up panel), lepton $g-2$ of 
 $\Delta a_\mu$  and $\Delta a_e$(left-down panel),  and  $\sum D_{\nu}-\delta_{\rm CP}$ in meV unit (right-down panel), respectively.
All the color legends and the line are the same as the ones in Fig.~\ref{fig:nhbdm1}.
These figures show that the predicted cLFV branching ratios and lepton $g-2$ are similar to the case of BDM, therefore it is difficult to test experimentally.

%%%%%%%%%%%%%%%%%%%
\begin{figure}[tb]\begin{center}
\includegraphics[width=77mm]{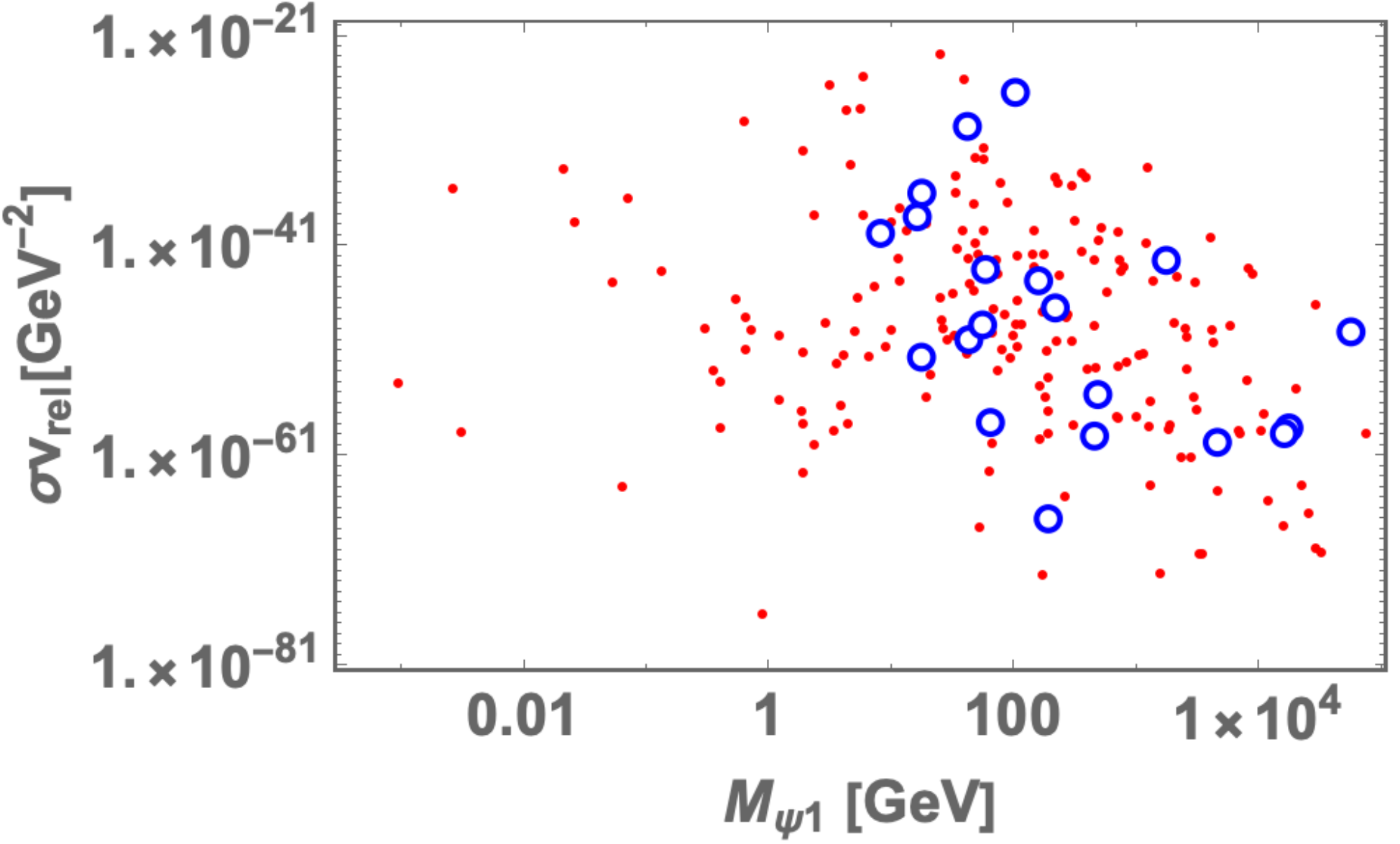}
\caption{Allowed region of the DM cross section to explain the relic density in terms of DM mass. 
All the color legends and the line are the same as the ones in Fig.~\ref{fig:nhbdm1}.   }
\label{fig:nhfdm3}
\end{center}\end{figure}
%%%%%%%%%%%%%%%%%%%
%
\noindent In Fig.~\ref{fig:nhfdm3}, we show the allowed region of the DM cross section to explain the relic density in terms of DM mass. 
All the color legends and the line are the same as the ones in Fig.~\ref{fig:nhbdm1}. This plot indicates that fermionic DM candidate is disfavored since the obtained cross section is much lower than the required cross section $\sim10^{-9}$ GeV$^{-2}$ to explain the relic density.

\subsection{IH}
\label{subsec:ih_num}
\noindent It was found that unlike NH, the IH scenario does not yield a unique solution unless the parameter space is constrained. Therefore, rather than performing a global analysis as done for NH, we present a specific benchmark (BP) for IH.

\subsubsection{Bosonic DM: $S$ state}
\noindent Table~\ref{tab:ihbdm} lists the BP that minimizes $\Delta \chi^2$, where our DM is fixed to be half of the SM Higgs mass $\sim63$ GeV and 
the resulting minimum value is approximately $0.200$.
The corresponding values for $\tilde g$ and $\tilde f$ are also provided as follows:
\begin{align}
& \tilde g \times 10^3
\approx 
 \left(\begin{array}{ccc} 
-0.890644 + 1.97046 i & -229.461 + 164.759 i& -1.46252 + 1.0415  i \\
114.175 - 22.8643 i & 1.34085 + 42.0388  i & -1326.2 - 1491.42  i \\
-209.66 - 876.698  i & -437.105 + 40.0709 i & 10^3  \end{array} \right),\\
& \tilde f \times 10^3
\approx 
 \left(\begin{array}{ccc} 
-1009.05 - 1888.65  i & -3.17909 - 3.85315  i& -1.63715 - 0.329819 i \\
3.26515 - 152.467 i & 171.427 - 88.4997 i & -29.766 - 109.796 i \\
-9.45895 - 4.10588 i & 25.7098 - 265.766 i & 10^3  \end{array} \right).
\label{eq:cgd-mtx}
\end{align}
{Consequently, the benchmark point simultaneously accommodates neutrino oscillation data and all phenomenological constraints. Since the bosonic singlet mass is fixed near the Higgs-resonance region, the singlet scalar remains a viable dark matter candidate in the IH case as well. }

%%%%%%%%%%%%%%%%%%
\begin{table}[tb]
    \setlength\tabcolsep{0.2cm}
    \begin{tabular}{c|c||c|c||c|c}
%        \toprule
\hline
        Parameter    &  BF & Parameter & BF & Parameter & BF \\ \hline \hline
          $\tilde M_{\psi_1}$ &  $164$ & $\tilde M_{\psi_2}$ &  $142$ & $M_{\psi_3}$/[GeV] &$92.7$ \\ \hline
           $\tilde m_{1}$ &  $0.676$ & $\tilde m_{2}$ &  $49.6$ & $s_\theta$ & $0.0273$ \\ \hline
       $g_{33}\times10^5$  & $-9.51+4.78 i$ &  $f_{33}\times10^7$  & $11.0+5.55 i$  &
 $\kappa_\nu^2$/[GeV$^2$] & $4.41\times10^{-24}$ \\ \hline
  $-$ & $-$ &
 $\Delta a_e$  &$5.59\times10^{-30}$ &  $\Delta a_\mu$ & $9.12\times10^{-27}$ \\ \hline
 BR($\mu\to e\gamma$)  &$1.87\times10^{-37}$ &
 BR($\tau\to e\gamma$)  &$1.35\times10^{-43}$ &  BR($\tau\to \mu\gamma$) & $1.66\times10^{-41}$ \\ \hline
 %%%
%$\beta_{\ell}/\alpha_\ell$ & $0.994$ & $\delta_{\ell}/\alpha_\ell$  & $0.000365$ & $\gamma_{\ell}/\alpha_\ell$ & $0.123$ \\ \hline
%%%
$s_{12}$ & 0.554 &$s_{23}$ & 0.752 &$s_{13}$ & 0.150 \\ \hline
$\Delta m^2_{\rm sol}$ & $7.52\times 10^{-5}~{\rm eV}^2$ &  $\Delta m^2_{\rm atm}$ & $2.51\times 10^{-3}~{\rm eV}^2$ &
$\delta_{\rm CP}$ & $29.5^{\circ}$ \\ \hline
$\sum D_{\nu}$ & $112~{\rm meV}$ & $- $ 
& $ -$ & $ -$ & $- $ \\
%        \bottomrule
\hline
    \end{tabular}
    \caption{\label{tab:ihbdm}%
     BF data set in case of bosonic DM in IH, where $\Delta\chi^2\approx0.200$.}
\end{table}

\subsubsection{Fermionic DM: $\psi_1$ state}
\noindent Table~\ref{tab:ihfdm} lists the BP that minimizes $\Delta \chi^2$.
The resulting minimum value is approximately $0.148$.
The corresponding values for $\tilde g$ and $\tilde f$ are also provided as follows:
\begin{align}
& \tilde g \times 10^3
\approx 
 \left(\begin{array}{ccc} 
75.3899 - 79.9135i & -1.10263 + 0.0792865 i& 0.0115587 - 0.0801839 i \\
-7.77716 + 26.6837i & -459.569 - 3588.96i & -2.9311 + 11.6813i \\
0.0249052 + 0.108523i & 8.89349 + 16.6716 i & 10^3  \end{array} \right),\\
& \tilde f \times 10^3
\approx 
 \left(\begin{array}{ccc} 
1832.83 + 2268.43 i & -97.301 + 383.696 i& -1.19209 + 1.64211 i \\
7.70775 - 0.953411 i & -26.8426 + 38.648 i & -900.194 - 499.272 i \\
30.3331 - 16.0167 i & 5.87214 + 38.3581 i & 10^3  \end{array} \right).
\label{eq:cgd-mtx}
\end{align}
As shown in this Table \ref{FDMBF}, the cross section required to account for the observed relic density of the FDM case is significantly smaller than the expected value of $\sim10^{-9}$ GeV$^{-2}$.
Consequently, the FDM cannot be the dominant component of DM in the IH case either. 

%%%%%%%%%%%%%%%%%%
\begin{table}[tb]
    \setlength\tabcolsep{0.2cm}
    \begin{tabular}{c|c||c|c||c|c}
%        \toprule
\hline
        Parameter    &  BF & Parameter & BF & Parameter & BF \\ \hline \hline
          $\tilde M_{\psi_1}$ &  $2.14\times10^{-5} $ & $\tilde M_{\psi_2}$ &  $0.961$ & $M_{\psi_3}$/[GeV] &$4.35\times10^3$ \\ \hline
           $\tilde m_{1}$ &  $2.41\times10^{-5} $ & $\tilde m_{2}$ &  $3.54$ & $s_\theta$ & $0.127$ \\ \hline
       $g_{33}\times10^4$  & $-3.85-1.05 i$ &  $f_{33}\times10^9$  & $-8.82+2.40 i$  &
 $\kappa_\nu^2$/[GeV$^2$] & $1.59\times10^{-22}$ \\ \hline
  $\frac{\langle\sigma v_{\rm rel}\rangle}{\rm GeV^{-2}}$ & $2.90\times10^{-35}$ &
 $\Delta a_e$  &$4.21\times10^{-34}$ &  $\Delta a_\mu$ & $1.97\times10^{-30}$ \\ \hline
 BR($\mu\to e\gamma$)  &$2.52\times10^{-48}$ &
 BR($\tau\to e\gamma$)  &$2.10\times10^{-50}$ &  BR($\tau\to \mu\gamma$) & $2.96\times10^{-50}$ \\ \hline
 %%%
%$\beta_{\ell}/\alpha_\ell$ & $0.994$ & $\delta_{\ell}/\alpha_\ell$  & $0.000365$ & $\gamma_{\ell}/\alpha_\ell$ & $0.123$ \\ \hline
%%%
$s_{12}$ & 0.553 &$s_{23}$ & 0.697 &$s_{13}$ & 0.150 \\ \hline
$\Delta m^2_{\rm sol}$ & $7.54\times 10^{-5}~{\rm eV}^2$ &  $\Delta m^2_{\rm atm}$ & $2.50\times 10^{-3}~{\rm eV}^2$ &
$\delta_{\rm CP}$ & $173^{\circ}$ \\ \hline
$\sum D_{\nu}$ & $99.2~{\rm meV}$ & $- $ 
& $ -$ & $ -$ & $- $ \\
%        \bottomrule
\hline
    \end{tabular}
    \caption{\label{tab:ihfdm}%
     BF data set in case of fermionic DM in IH, where $\Delta\chi^2\approx0.148$.}
     \label{FDMBF}
\end{table}

\section{Conclusion and discussion}
\noindent In this paper, we have constructed a radiative Dirac neutrino mass model based on a non-invertible fusion rule derived from $Z_3$ gauging of $Z_3\times Z'_3$ fusion rule.
The proposed framework forbids tree-level Yukawa couplings and ensures that neutrino masses arise only at the one-loop level, thereby realizing a minimal Dirac seesaw mechanism. The model introduces vector-like neutral fermions and inert scalar fields, which simultaneously generate neutrino masses and provide a viable DM candidate.
Phenomenological analyses demonstrate that:
\begin{enumerate} 
\item The model is consistent with current neutrino oscillation data, including NuFit 6.1 and JUNO constraints.
\item Lepton flavor violating processes such as $\mu\to e\gamma$ remain well below experimental bounds.
\item Contributions to the lepton anomalous magnetic moments are small, ensuring compatibility with precision measurements.
\item The bosonic singlet $S$ emerges as a promising DM candidate, while the fermionic option is strongly disfavored due to insufficient annihilation cross sections..
\end{enumerate}

\noindent The significance of this work lies in demonstrating that non-invertible fusion rules can serve as a powerful organizing principle for constructing minimal and phenomenologically viable extensions of the Standard Model. By forbidding Majorana mass terms while allowing loop-induced Dirac masses, the framework provides a natural explanation for the smallness of neutrino masses and establishes a direct connection to DM physics.
Although the model successfully satisfies all current experimental constraints, its testability remains limited. The predicted branching ratios for charged lepton flavor violation and the contributions to lepton $g-2$
 are far below present sensitivities, making near-term experimental verification challenging. Nevertheless, the bosonic DM candidate offers a concrete scenario that could be probed through precision cosmology and collider searches.
Future directions include exploring collider signatures of the inert scalar and exotic fermions, investigating possible connections to baryogenesis, and examining alternative fusion rules. These extensions may broaden the phenomenological implications and provide new avenues for experimental tests.

\noindent In conclusion, the proposed Dirac one-loop seesaw model with non-invertible fusion rules offers a consistent and minimal framework that simultaneously addresses neutrino masses and DM, while remaining fully compatible with current experimental data. It establishes a foundation for further theoretical developments and motivates future experimental efforts to probe the underlying symmetry structures beyond the Standard Model.

\begin{acknowledgments}
\noindent H.O. is supported by the Henan Provincial Overseas High-level Talent Recruitment Plan and the Zhongyuan Talent Plan.
\end{acknowledgments}

% Including title
%\bibliographystyle{utphys}
\bibliography{references.bib}
\end{document}